\documentclass[aps,prc,twocolumn,10pt, superscriptaddress, showpacs, floatfix]{revtex4-1}

\usepackage{amssymb,epsfig}
\usepackage{bbold}

\newcommand{\be}{\begin{equation}}
\newcommand{\ee}{\end{equation}}
\newcommand{\bea}{\begin{eqnarray}}
\newcommand{\eea}{\end{eqnarray}}

\begin{document}

\title{Fermionic equations of motion in strongly-correlated media: applications to the nuclear many-body problem}
\author{Elena Litvinova}
\affiliation{Department of Physics, Western Michigan University, Kalamazoo, MI 49008, USA}
\affiliation{Facility for Rare Isotope Beams, Michigan State University, East Lansing, MI 48824, USA}
\affiliation{GANIL, CEA/DRF-CNRS/IN2P3, F-14076 Caen, France}

\date{\today}

\begin{abstract}
These notes summarise the lectures given at the International School of Physics 'Enrico Fermi' in Summer 2024 in Varenna (Italy) about the strongly coupled quantum many-body theory and its applications to nuclear structure. The lectures present a rather short overview of the subject with an emphasis on the analytical aspects of the nuclear many-body problem, aiming at a deep understanding of the complexity of strongly coupled nucleonic states and emergent collective phenomena.  The major pedagogical focus is recognizing how all the models describing nuclear dynamics follow from a unified model-independent framework formulated in the universal language of quantum field theory. In particular, connections between the classes of ab initio, density functional theory, and beyond mean-field approaches are made accessible. Approximations of varying complexity are discussed in applications to excited states of medium-heavy nuclei.

\end{abstract}

\maketitle

\section{Introduction} 

The nuclear many-body problem underlies nearly all the physics frontiers and applications, from fundamentals to technologies. The nuclear structure theory demonstrated impressive analytical and computational developments over the years; however, accurate quantitative predictions of nuclear spectral properties remain challenging. 
Quantifying nuclear structure by means of the Green functions (GF) is one of the most universal and convenient methods since the Green functions, or propagators, belong to a general class of correlation functions (CFs) bridging the physics of strongly correlated quantum systems across the energy scales. The direct relationships of the GFs to the most accessible observables, in particular, in the many-body fermionic systems, are widely acknowledged \cite{Matsubara1955,Watson1956,Brueckner1955,Brueckner1955a,Martin1959,Ethofer1969,SchuckEthofer1973} and help establish patterns and emergent phenomena with common features for systems of various origins. 
For instance, the single-nucleon propagators in atomic nuclei characterize the energies of odd-particle systems and their orbital occupancies which can be benchmarked by transfer or knock-out reactions. 
The nuclear response to the most accessible experimental probes is associated with the two-time particle-hole, or two-quasiparticle, propagators. Superfluid effects can be encoded in the fermionic pair propagators and benchmarked by the nucleonic pair (e.g., deuteron) transfer, and the residues of those propagators are related to the pairing energy gaps \cite{Gorkov1958,Kadanoff1961}. 

The higher-rank, e.g., three and more, fermionic CFs enter the theory via the interaction kernels of the equations of motion (EOMs) for the low-rank propagators \cite{Martin1959,Ethofer1969,EthoferSchuck1969,SchuckEthofer1973}. 
Generally, for a given GF, the dynamical component of its EOM's interaction kernel contains a CF of a higher rank than the rank of the reference propagator. The higher-rank propagators correspond to correlated multi-fermion structures, or configurations, embedded in the medium and are responsible for the dynamical effects of long-range correlations while giving feedback on the short-range physics via the self-consistent EOMs. In these lectures, we will see that in the intermediate and strong coupling regimes, these higher-rank GFs encode emergent collective effects. At intermediate coupling associated with atomic nuclei, identifying the emergent degrees of freedom helps establish the order parameters relevant to the correlated medium, and the connections between the elementary and emergent degrees of freedom promote the theory across the energy scales. 

Most often, textbook formulations of the GF method operate non-symmetric dynamical kernels \cite{Migdal1967}, which are subsequently expanded in perturbation series or factorization in terms of single-fermion GFs. 
Peter Schuck and coauthors emphasized the important benefit of bringing the two-time dynamical kernels to a symmetric form  \cite{AdachiSchuck1989,Danielewicz1994, DukelskyRoepkeSchuck1998} for finding non-perturbative approximations to the interaction kernels via 
cluster decompositions. Examples include
retaining formally exact two-fermion CFs, with applications ranging from particle physics \cite{Popovici2010,Popovici2011} to nuclear structure \cite{LitvinovaSchuck2019,Litvinova2023a} to quantum chemistry and condensed matter physics \cite{Tiago2008,Martinez2010,Sangalli2011,SchuckTohyama2016,Olevano2018,Schuck2021}.
Interestingly, in this formalism, beyond-mean-field (BMF) approaches based on effective interactions and widely used in nuclear structure computation can be linked to the EOMs derived from Hamiltonians with bare fermionic interactions \cite{LitvinovaSchuck2019,Schuck2019,Schuck2021,LitvinovaSchuck2020,Litvinova2021}.  
The BMF approaches or those going beyond the random phase approximation (RPA) that account for emergent collective effects of the nuclear medium are of particular interest for medium-heavy nuclei with pronounced collectivity. They can be continuously derived ab initio, i.e., employing only the bare interaction, by retaining the correlated pairs of fermionic quasiparticles in the dynamical kernels of the EOMs for the propagators related to the required observables. These correlated pairs are known as phonons, which emerge as mediators of the dynamical in-medium interaction between fermions and exist on shells as collective excitations. The interaction, or coupling, between quasiparticles and phonons is the central part of the  
nuclear field theory (NFT) elaborated by the Copenhagen-Milano school, particularly Ricardo Broglia \cite{BohrMottelson1969,BohrMottelson1975,Broglia1976,BortignonBrogliaBesEtAl1977,BertschBortignonBroglia1983,Barranco2017}, NFT variants based on the Migdal's theory \cite{KamerdzhievSpethTertychny2004,Tselyaev1989,LitvinovaTselyaev2007} and the quasiparticle-phonon model (QPM) of V.G. Soloviev and collaborators \cite{Soloviev1992,Malov1976}.

By definition of the ab initio EOM, the static and dynamical kernels of the two-quasiparticle CFs play the role of the in-medium interaction between the fermions. The former kernel is responsible for the short-range correlations, while the latter governs the long-range ones.
As will be discussed in detail below, the short-range kernel contains contractions of the bare fermionic interaction with two-body densities, and the long-range one accommodates fully correlated four-fermion CFs doubly contracted with the bare interaction. The presence of these CFs in the interaction kernels makes the EOM non-linear and the exact solution intractable, but even for the approximations with reasonably factorized dynamical kernels, entering the self-consistent cycle of the non-linear EOM requires some educated guess about the static kernel. While the fully consistent ab initio calculations are not yet available, in practical applications, the static kernel can be well approximated by effective interactions.  The latter do not rely on the bare nucleon-nucleon (NN) forces but instead are fitted to properties of finite nuclei, for example, their masses and radii, as it is done in the density functional theories (DFTs).
In this approach, the static kernel of the, e.g., two-fermion EOM, overlaps with the implicitly included dynamical one, which, however, can be corrected by subtraction of the dynamical kernel at zero transition frequency recovering the static limit of the full kernel as proposed in Ref. \cite{Tselyaev2013}. This allows one to avoid double countings associated with the use of effective interactions in the self-consistent implementations of the response theory \cite{LitvinovaRingTselyaev2007,LitvinovaRingTselyaev2008,LitvinovaRingTselyaev2010,LitvinovaRingTselyaev2013,Tselyaev2018,LitvinovaSchuck2019,Litvinova2023}. 
    
As the majority of atomic nuclei and many other fermionic systems are essentially superfluid, an EOM for the single-quasiparticle propagator has been formulated for the superfluid case in Ref. \cite{Litvinova2021a} continuing the research line of the preceding work \cite{VanderSluys1993,Avdeenkov1999,Avdeenkov1999a,BarrancoBrogliaGoriEtAl1999,BarrancoBortignonBrogliaEtAl2005,Tselyaev2007,LitvinovaAfanasjev2011,Litvinova2012,AfanasjevLitvinova2015,IdiniPotelBarrancoEtAl2015,Soma2011,Soma2013,Soma2014a,Soma2021}. In Ref. \cite{Litvinova2021a}, the superfluid pairing correlations were included in the single-fermion self-energy derived from the bare Hamiltonian keeping the 2$\times$2 matrix structure of the Hartree-Fock-Bogolyubov (HFB) solutions. Applying the cluster decomposition to its dynamical part in a symmetric form and retaining the two-fermion CFs responsible for the leading emergent collective effects led to a formal extension of Gor'kov's theory of superfluidity. The transformation of the 2$\times$2 single-fermion EOM to the HFB  basis enabled a natural unification of the normal and pairing phonon modes under the {\it average} particle number conservation condition. This also leads to a
remarkable compactification of the Dyson equation and much more efficient handling of the fermionic self-energy beyond the HFB approach. The extraction of two-fermion CFs contracted with the interaction matrix elements was used to establish a mapping of the factorized dynamical kernel to the quasiparticle-vibration coupling (qPVC). 
A similar strategy was implemented for the ab initio EOM for the response function, which has been worked out in the HFB basis from the start for applications to superfluid fermionic systems with strong coupling \cite{Litvinova2022}. This allowed for the formulation of dynamical kernels with varying correlation content also for such systems, reproducing the limits of the known BMF approaches, such as the second RPA (SRPA), NFT, and the extended Migdal's theory, and paved the way to the more complex interaction kernels which are needed for accurate spectral calculations of open-shell nuclei. 

The formal part of these lecture notes is focused on the EOMs for the two-point fermionic propagators in strongly correlated media with an emphasis on the dynamical interaction kernels. Starting with the many-body Hamiltonian confined by a two-body interaction between two fermions in the vacuum,  it is shown by continuous derivation how the EOMs for the two-point in-medium fermionic propagators acquire the Dyson form.  I elaborate specifically on the one-fermion and two-fermion CFs related to the single-particle observables and response to external probes of various kinds. We will see that, before taking any approximation, the interaction kernels of the respective EOMs decompose into static and dynamic (time-dependent) contributions.  The latter translates to the energy-dependent, and the former maps to the energy-independent terms in the energy domain. Furthermore, it is argued that the static kernels are not well-known due to the difficulty of their accurate determination, which justifies the efficiency of applying the concept of effective interaction compared to fully ab initio approaches operating exclusively on bare interactions. I dwell particularly on the dynamic terms, which generate long-range correlations while giving feedback on their short-range static counterparts. The origin, forms, and various approximations for the dynamical kernels of one-fermion and two-fermion propagators, most relevant in the intermediate-coupling regime, are discussed.

Special emphasis
is put on the aspects elaborated and inspired by the scientific work of Ricardo Broglia. This pertains to the qPVC approaches under NFT progressed over decades \cite{Bes1975,BesBrogliaDusselEtAl1976,BortignonBrogliaBesEtAl1977,BertschBortignonBroglia1983,BarrancoBrogliaGoriEtAl1999,Terasaki2002,BarrancoBortignonBrogliaEtAl2005,NiuColoVigezziEtAl2014,NiuNiuColoEtAl2015,Niu2016,Barranco2017}. The idea of qPVC is a critical ingredient for the description of medium-mass and heavy nuclei which offers an optimal combination of accuracy and feasibility accessible by making use of modern effective interactions self-consistently and the qPVC as the leading approximation to emergent collectivity \cite{LitvinovaRingTselyaev2008,LitvinovaRingTselyaev2010,Gambacurta2011,Gambacurta2015,NiuColoVigezziEtAl2014,NiuNiuColoEtAl2015,RobinLitvinova2016,Tselyaev2018,Robin2019}. 
These studies are dominated by the dynamical kernels of two-particle-two-hole ($2p2h$) configuration complexity beyond the simplistic RPA. After decades of exploring such approaches, it has become evident that, although the latter configuration complexity is necessary to improve the description of nuclear excited states, they are still insufficient for reproducing the observed spectral richness and spectroscopically accurate results.   


A path to higher configuration complexity was set by the QPM of V.G. Soloviev and collaborators \cite{Soloviev1992,Soloviev:1977fmy,Ponomarev2001}. QPM is formulated in the basis of phonons evaluated by the quasiparticle random phase approximation (QRPA), i.e., correlated two-quasiparticle pairs, and admits complex wave functions in the form of multiphonon configurations \cite{Andreozzi2008,Knapp:2014xja,Bacca2014,Knapp:2015wpt,DeGregorio2016,DeGregorio2016a}. 
QPM implementations were advanced to the $3p3h$, or three-phonon, configuration complexity for medium-heavy nuclei in model spaces limited by low energy and showed success \cite{Ponomarev1999,LoIudice2012,Savran2011,Tsoneva2019,Lenske:2019ubp}. 
Although these implementations exist only in non-selfconsistent frameworks and require adjustments of the interaction parameters to data for each multipole, they show unambiguously that to reproduce the richness of the observed spectra, at least $3p3h$ configuration complexity should be included in the theory. Therefore, considerable effort was dedicated to including such configurations, which has become possible recently within a relativistic framework in larger model spaces up to high energy (25-30 MeV) \cite{LitvinovaSchuck2019,Litvinova2023a,Muescher2024,Novak2024} with the current computational capabilities. 
In light of the high demand from many frontier applications, it is highly desirable to formulate and implement a consistent, accurate, and predictive approach valid across the entire nuclear chart or the major part of it. To achieve this goal, the nuclear theory should be advanced in its two major components, essentially interrelated: (i) the NN interactions, both bare and effective, and (ii) the advanced solutions to the many-body problem. In these notes, I focus on the latter solutions aiming at an accurate modeling of nucleonic dynamics in a correlated medium using an unspecified NN interaction between two fermions in the vacuum as an input. The formalism section closely follows that of Ref. \cite{Litvinova2023a} complemented by extended methodical recommendations.

\section{Fermionic propagators in a correlated medium}
\label{Propagators}
\subsection{Definitions and microscopic input to the many-body theory}

 The starting point for the many-body theory can be the system's Lagrangian or, alternatively, its Hamiltonian. 
The Hamiltonian formulation is especially convenient for dynamical theory, which tracks explicit time dependence.
The many-body Hamiltonian in the field-theoretical representation generally reads
\be
H = H^{(1)} + V^{(2)} + W^{(3)} + ... \ ,
\label{Hamiltonian}
\ee
where the operator $H^{(1)}$ describes the one-body contribution:
\be
H^{(1)} = \sum_{12} t_{12} \psi^{\dag}_1\psi_2 + \sum_{12}v^{(MF)}_{12}\psi^{\dag}_1\psi_2 \equiv \sum_{12}h_{12}\psi^{\dag}_1\psi_2
\label{Hamiltonian1}
\ee
and $h_{12}$ are the matrix elements containing the kinetic energy $t$ and the mean-field $v^{(MF)}$ part of the interaction, in case the latter mean field is present. The operators  $\psi_1, \psi^{\dagger}_1$ are the destruction and creation fermionic field operators in some basis completely characterized by the number indices. In this work, we focus on the EOMs for fermionic fields, while bosonic fields will mediate the interaction between fermions. The latter is quantified by   the interaction operator $V^{(2)}$ between two-fermions:
\be
V^{(2)} = \frac{1}{4}\sum\limits_{1234}{\bar v}_{1234}{\psi^{\dagger}}_1{\psi^{\dagger}}_2\psi_4\psi_3,
\label{Hamiltonian2}
\ee
where the antisymmetrized matrix elements ${\bar v}_{1234} = {v}_{1234} - {v}_{1243}$ express the meson exchange either explicitly or in some form of effective field theory. 
$W^{(3)}$ formally stands for the three-body forces, which are neglected in the presented formalism because the numerical implementation of the theory discussed in Section \ref{calculations} is performed in a relativistic framework. 
We will not use explicitly covariant notations in these lectures; however, bear in mind that the relativistic field-theoretical Hamiltonian of interacting nucleons is defined by the terms having the same nucleonic field operator structure as $H^{(1)}$ and $V^{(2)}$ \cite{Brockmann1978,Boyussy1987}. Because of that, the general structure of the fermionic EOMs defined by the fermionic field operator composition of $H$ remains the same. 
Our expressions for the one-fermion EOM with a non-symmetric form of the dynamical kernel can be compared, for instance, with those of Refs. \cite{Poschenrieder1988, Poschenrieder1988a}. 
A justification for neglecting $W^{(3)}$ is the considerably smaller role of three-body forces in a relativistic theory than in a non-relativistic one. Whether the relativistic three-body forces should be included in the description of nuclear systems is not completely clear. Quantitative studies of this subject were reported only for few-body systems \cite{Danielewicz1979,Karmanov2011}. An adequate non-perturbative description of in-medium CFs should generate the three- and higher-body forces identifiable as in Ref. \cite{Karmanov2011} and therefore realistically incorporate their effects, at least in medium-mass and heavy nuclei. 

The operators of fermionic (nucleonic) fields form the usual anticommutator algebra
\bea
[\psi_1,{\psi^{\dagger}}_{1'}]_+ \equiv \psi_1{\psi^{\dagger}}_{1'}  +  {\psi^{\dagger}}_{1'}\psi_1 = \delta_{11'}, \nonumber \\
\left[ \psi_1,{\psi}_{1'} \right]_{+}  = \left[ {\psi^{\dagger}}_1,{\psi^{\dagger}}_{1'}\right]_+ = 0,
\label{anticomm}
\eea
and undergo time evolution in the Heisenberg picture:
\be
\psi(1) = e^{iHt_1}\psi_1e^{-iHt_1}, \ \ \ \ \ \ {\psi^{\dagger}}(1) = e^{iHt_1}{\psi^{\dagger}}_1e^{-iHt_1}.
\ee

The CF of two fermionic field operators introduces the one-fermion in-medium propagator or its real-time Green function:
\be
G(1,1') \equiv G_{11'}(t-t') = -i\langle T \psi(1){\psi^{\dagger}}(1') \rangle,
\label{spgf}
\ee
via the chronological ordering operator $T$ and the averaging $\langle ... \rangle$ over the many-body ground state of the system containing $N$ interacting fermions (e.g., nucleons). The correlation function $G_{11'}(t-t')$ describes the probability amplitude of a single fermion to travel through a medium of identical interacting fermions.

Basis choice is essential as it can simplify applications of formalism considerably. On the contrary, an inconveniently chosen basis can greatly complicate calculations.
One common choice in the EOM method is the basis of fermionic states $\{1\}$
diagonalizing simultaneously the one-body  part of the Hamiltonian $H^{(1)}$: $h_{12} =  \delta_{12}\varepsilon_1$ and the corresponding density matrix. 
As follows from Eq. (\ref{spgf}), the propagator $G_{11'}(t-t')$ is time translational invariant, i.e., depends explicitly on the time difference $\tau = t-t'$. Thus, its Fourier image is a function of a single energy (frequency). Indeed, inserting the operator $\mathbb{1} = \sum_n |n\rangle\langle n|$ with the complete set of the many-body states $|n\rangle$, one arrives at the spectral expansion, or K\"all\'en - Lehmann representation \cite{Kallen1952,Lehmann1954}:
\bea
G_{11'}(\varepsilon) = \sum\limits_{n}\frac{\eta^{n}_{1}\eta^{n\ast}_{1'}}{\varepsilon - (E^{(N+1)}_{n} - E^{(N)}_0)+i\delta} +  \nonumber \\
+ \sum\limits_{m}\frac{\chi^{m}_{1}\chi^{ m\ast}_{1'}}{\varepsilon + (E^{(N-1)}_{m} - E^{(N)}_0)-i\delta}. 
\label{spgfspec}
\eea
Thus, in the energy domain, the propagator (\ref{spgf}) is a series of simple poles with factorized residues. This is the common feature of propagators in quantum field theory (QFT), expressing its locality and unitarity. The poles are the formally exact energies $E^{(N+1)}_{n} - E^{(N)}_0$ and $-(E^{(N-1)}_{m} - E^{(N)}_0)$ of the  $(N+1)$-particle and $(N-1)$-particle systems, respectively, above the ground state of the reference $N$-particle system.
The residues corresponding to these poles are the following matrix elements 
\be
\eta^{n}_{1} = \langle 0^{(N)}|\psi_1|n^{(N+1)} \rangle , \ \ \ \ \ \ \ \  \chi^{m}_{1} = \langle m^{(N-1)}|\psi_1|0^{(N)} \rangle 
\label{etachi}
\ee
of the fermionic operators between the ground state $|0^{(N)}\rangle$ of the $N$-particle system and states $|n^{(N+1)} \rangle$ and $|m^{(N-1)} \rangle$ of the neighboring systems. The latter are the overlaps of the single-particle (single-hole) configurations $\{1\}$ on top of the ground state $|0^{(N)}\rangle$ in the $n$-th ($m$-th) state of the $(N+1)$-particle  ($(N-1)$-particle) systems. The residues in Eq. (\ref{spgfspec}) are thus associated with the (generally fractional) occupancies of the correlated fermionic states.

The EOM for the propagator $G_{11'}(t-t')$, introduced and discussed in the next subsection, connects it to higher-rank propagators via interaction kernels. The most relevant are the two-time, or two-point, CFs: the particle-hole propagator, or response function, and the particle-particle, or fermionic pair, propagator. As we will see in the following, these CFs figure in the leading beyond-mean-field contributions to the fermionic self-energy, introducing the coupling of single-particle motion to collective degrees of freedom. The latter ones emerge as correlated (quasi)particle pairs forming boson-like structures exchanged between fermions. This phenomenon is known as(quasi)particle-vibration coupling, or (quasi)particle-phonon coupling, while the quasi-bosons, or phonons, associated with the response and pair CFs, underly the zero sound and superfluidity phenomena, respectively. Besides their in-medium realm, the phonons also exist on-shell in the form of collective normal and pairing vibrations, which can be actualized by electromagnetic, hadronic, and weak probes or by nucleon pair transfer in the case of pairing.

The particle-hole propagator, or response, is defined as:
\bea
R(12,1'2') &\equiv& R_{12,1'2'}(t-t') = -i\langle T\psi^{\dagger}(1)\psi(2)\psi^{\dagger}(2')\psi(1')\rangle \nonumber \\ &=&
-i\langle T(\psi^{\dagger}_1\psi_2)(t)(\psi^{\dagger}_{2'}\psi_{1'})(t')\rangle,
\label{phresp}
\eea
and the pair propagator reads:
\bea
G(12,1'2') &\equiv& G_{12,1'2'}(t-t') = -i\langle T \psi(1)\psi(2){\psi^{\dagger}}(2'){\psi^{\dagger}}(1')\rangle \nonumber \\ &=& 
-i\langle T(\psi_1\psi_2)(t)(\psi^{\dagger}_{2'}\psi^{\dagger}_{1'})(t')\rangle.
\label{ppGF} 
\eea
Both CFs, like the single-particle propagator, depend on a single time difference, and it is implied that $t_1 = t_2 = t, t_{1'} = t_{2'} = t'$.

After inserting the completeness relation between the operator pairs and performing the Fourier transformations, analogously to the one-fermion case, the spectral expansions of CFs (\ref{phresp},\ref{ppGF}) turn out to be the following series:
\be
R_{12,1'2'}(\omega) = \sum\limits_{\nu>0}\Bigl[ \frac{\rho^{\nu}_{21}\rho^{\nu\ast}_{2'1'}}{\omega - \omega_{\nu} + i\delta} -  \frac{\rho^{\nu\ast}_{12}\rho^{\nu}_{1'2'}}{\omega + \omega_{\nu} - i\delta}\Bigr]
\label{respspec}
\ee
\be
G_{12,1'2'}(\omega) =  \sum\limits_{\mu} \frac{\alpha^{\mu}_{21}\alpha^{\mu\ast}_{2'1'}}{\omega - \omega_{\mu}^{(++)}+i\delta} - \sum\limits_{\varkappa} \frac{\beta^{\varkappa\ast}_{12}\beta^{\varkappa}_{1'2'}}{\omega + \omega_{\varkappa}^{(--)}-i\delta},
\label{resppp}
\ee
similarly local and unitary, as higher operator rank two-point CFs. The poles are located at the energies $\omega_{\nu} = E_{\nu} - E_0, \omega_{\mu}^{(++)}= E_{\mu}^{(N+2)} - E_0^{(N)}$, and $\omega_{\mu}^{(--)}= E_{\varkappa}^{(N-2)} - E_0^{(N)}$ of the systems with $N$ and $N\pm 2$ particles, respectively. The sums in Eqs. (\ref{spgfspec},\ref{respspec},\ref{resppp}) are formally complete, i.e., include both the discrete and continuum states. 

The residues are products of the matrix elements
\bea
\rho^{\nu}_{12} = \langle 0|\psi^{\dagger}_2\psi_1|\nu \rangle \ \ \ \ \ \ \ \ \ \ \ \ \ \ \ \ \ \  \\
\label{trden}
\alpha_{12}^{\mu} = \langle 0^{(N)} | \psi_2\psi_1|\mu^{(N+2)} \rangle  \ \ \ \ \ \ \beta_{12}^{\varkappa} = \langle 0^{(N)} | \psi^{\dagger}_2\psi^{\dagger}_1|\varkappa^{(N-2)} \rangle \nonumber \\
\label{alphabeta}
\eea
are the transition densities of the normal $\rho^{\nu}_{12}$ and anomalous (pairing) $\alpha_{12}^{\mu}, \beta_{12}^{\varkappa}$ character. They are commonly interpreted as the weights of the particle-hole, two-particle, and two-hole configurations on top of the ground state $|0^{(N)}\rangle$ in the correlated excited states of the systems with matching particle numbers. The transition densities characterize probabilities of transitions between, e.g., $|0\rangle$ and $|\nu\rangle$, pair transfer amplitudes, and superfluid pairing spectral gaps \cite{LitvinovaSchuck2020}.
We note here that Eqs. (\ref{spgfspec},\ref{respspec} -- \ref{alphabeta}) are, in principle, exact, i.e., can be adapted for any physical approximations to the many-body states $|n\rangle, |m\rangle$, $|\nu\rangle$, $|\mu\rangle$, and $|\varkappa\rangle$. 

\subsection{Equation of motion for the single-fermion propagator: the normal phase}
\label{EOM1}

The canonical way of generating the EOM for the fermionic propagator (\ref{spgf}) in the Hamiltonian formalism is taking time derivatives with respect to the time variables. Here, we navigate the reader through the major steps, while the detailed derivation can be found in Refs. \cite{LitvinovaSchuck2019,Litvinova2021a}, in agreement with earlier work \cite{SchuckEthofer1973,Schuck1976,AdachiSchuck1989,Danielewicz1994,DukelskyRoepkeSchuck1998,Dickhoff2004,Dickhoff2005}. 

Differentiating Eq. (\ref{spgf}) with respect to $t$ gives:
\bea
\partial_t G_{11'}(t-t') = -i\delta(t-t')\langle [\psi_1(t),{\psi^{\dagger}}_{1'}(t')]_+\rangle + \nonumber \\
+ \langle T[H,\psi_1](t){\psi^{\dagger}}_{1'}(t')\rangle, \nonumber\\ 
\label{dtG}                           
\eea
where we denote, to simplify the notations, the time-dependent operator products and, in particular, the commutator as
$[H,\psi_1](t) = e^{iHt}[H,\psi_1]e^{-iHt}.$
First, the commutator with the one-body part of the Hamiltonian is evaluated explicitly. Then, isolating $G_{11'}(t-t')$, one obtains the equation:
\be
(i\partial_t - \varepsilon_1)G_{11'}(t-t') = \delta_{11'}\delta(t-t') + i\langle T[V,\psi_1](t){\psi^{\dagger}}_{1'}(t')\rangle.
\label{spEOM}
\ee
Second, the commutator with the interacting two-body part is evaluated, which leads to the first EOM, or EOM1:
\bea
(i\partial_t - \varepsilon_1)G_{11'}(t-t') &=& \delta_{11'}\delta(t-t') \nonumber \\
&+& 
\frac{i}{2}\sum\limits_{ikl}{\bar v}_{i1kl}
\langle T(\psi^{\dagger}_i\psi_l\psi_k)(t){\psi^{\dagger}}_{1'}(t')\rangle, \nonumber \\
\label{spEOM1}
\eea
where the Latin dummy indices have the same meaning as the number indices and mark the intermediate fermionic states in the working single-particle basis. Eq. (\ref{spEOM1}) is the most common form of the single-particle propagator EOM \cite{SchuckEthofer1973,Dickhoff2005,Poschenrieder1988}. The two-fermion CF on the right-hand side of Eq. (\ref{spEOM1}) is the heart of the dynamical interaction kernel which signals that the one-fermion propagator and the associated properties are coupled to higher-rank propagators. An EOM for this two-fermion CF can be generated, but its dynamical kernel contains a further higher-rank CF forming a complex hierarchy which is the characteristic feature of strongly-correlated systems described in the QFT language. In weakly coupled regimes, i.e., when the interaction $V$ contains a small parameter, the perturbation theory is a viable solution that has the advantage of controllability and quantifiable uncertainties, however, perturbative expansions do not converge at strong coupling. 

The latter is the case of nuclear systems, therefore, we focus on non-perturbative solutions in this discourse.
The Fourier image of Eq. (\ref{spEOM1}) 
reads
\be
G_{11'}(\omega) = G^{0}_{11'}(\omega) + \frac{1}{2}\sum\limits_{2ikl}G^{0}_{12}(\omega){\bar v}_{2ikl}G^{(2)}_{ilk,1'}(\omega),
\label{spEOM1a}
\ee
where we introduced $G^{0}_{11'}(\omega) = \delta_{11'}/(\omega - \varepsilon_1)$, which plays the role of the free, or uncorrelated, fermionic propagator, and the CF $G^{(2)}_{ilk,1'}$ on the right-hand side is the Fourier image of
\be
G^{(2)}_{ilk,1'}(t-t') = -i\langle T(\psi^{\dagger}_i\psi_l\psi_k)(t){\psi^{\dagger}}_{1'}(t')\rangle .
\label{G31}
\ee
Eq. (\ref{spEOM1a}) can be further transformed to the Dyson form \cite{Dickhoff2004,Dickhoff2005}. There exist various approximations to the integral part of the Eq. (\ref{spEOM1a}), specifically to the CF $G^{(2)}$.  The relativistic "$\Lambda^{00},\Lambda^{10}$", and "$\Lambda^{11} $" approximations are known from, e.g., Ref. \cite{Poschenrieder1988a}. They are based on factorizing the CF (\ref{G31}) into two one-fermion CFs (\ref{spgf}) of correlated or uncorrelated character. Another common approach is known as the Gor'kov theory of superfluidity which retains, in addition, one-fermion CFs with the same kind of field operators (anomalous Green functions) in the one-fermion CFs \cite{Gorkov1958,KucharekRing1991}. The detailed derivation of the Gor'kov theory from Eq. (\ref{spEOM1a}) can be found in Ref. \cite{Litvinova2021a}.

Before elaborating on feasible non-perturbative solutions of Eq. (\ref{spEOM1}), we transform it into a symmetric form, which provides 
more insights into the interacting part of the one-fermion EOM. This form is obtained by differentiating the last term on the right-hand side of Eq. (\ref{spEOM}) $R_{11'}(t-t') =  i\langle T[V,\psi_1](t){\psi^{\dagger}}_{1'}(t')\rangle$ with respect to $t'$: 
\bea
R_{11'}(t-t')\overleftarrow{\partial_{t'}} 
&=& -i\delta(t-t') \langle \bigl[[V,\psi_1](t),{\psi^{\dagger}}_{1'}(t')\bigr]_+\rangle - \nonumber \\ 
&-& \langle T[V,\psi_1](t)[H,{\psi^{\dagger}}_{1'}](t')\rangle,
\eea
which leads to the EOM2:
\bea
R_{11'}(t-t')(-i\overleftarrow{\partial_{t'}}  - \varepsilon_{1'}) &=& -\delta(t-t')\langle \bigl[ [V,\psi_1](t),{\psi^{\dagger}}_{1'}(t')\bigr]_+\rangle\nonumber \\
&+& i\langle T [V,\psi_1](t)[V,{\psi^{\dagger}}_{1'}](t')\rangle.
\label{EOMR}
\eea
Acting on the EOM1 (\ref{spEOM}) by the operator $(-i\overleftarrow{\partial_{t'}}  - \varepsilon_{1'})$ and performing the Fourier transformation, one obtains:
\be
G_{11'}(\omega) 
= G^{0}_{11'}(\omega) + 
\sum\limits_{22'}G^{0}_{12}(\omega)T_{22'}(\omega)G^{0}_{2'1'}(\omega),
\label{spEOM3}
\ee
with $T(\omega)$ being the Fourier image of the time-dependent $T$-matrix:
\bea
T_{11'}(t-t') &=& T^{0}_{11'}(t-t') + T^{r}_{11'}(t-t'), \nonumber\\
T^{0}_{11'}(t-t') &=& -\delta(t-t')\langle \bigl[ [V,\psi_1](t),{\psi^{\dagger}}_{1'}(t')\bigr]_+\rangle, \nonumber \\ 
T^{r}_{11'}(t-t') &=&  i\langle T [V,\psi_1](t)[V,{\psi^{\dagger}}_{1'}](t')\rangle.
\label{Toperator}
\eea
According to Eq. (\ref{spEOM3}), the complete in-medium one-fermion propagator $G$ is expressed via the free propagator $G^0$ and the $T$-matrix (\ref{Toperator}), which includes all possible interaction processes of a single fermion with the correlated medium. It is fundamentally split into the static, or instantaneous, part $T^{0}$ and dynamical, or time-dependent, part $T^r$ containing retardation effects. 
The EOM (\ref{spEOM3}) is more convenient in the Dyson form operating the irreducible part of the $T$-matrix with respect to the uncorrelated one-fermion propagator $G^{0}$. The irreducible part of the $T$-matrix is the self-energy (called also interaction kernel): $\Sigma = T^{irr}$ defined via:
\be
T(\omega) = \Sigma(\omega) + \Sigma(\omega) G^{0}(\omega)T(\omega).
\label{DysonT}
\ee
Combining Eq. (\ref{spEOM3}) in the operator form
\be
G(\omega) = G^{0}(\omega) + G^{0}(\omega)T(\omega)G^{0}(\omega)
\label{spEOM4}
\ee
and (\ref{DysonT}), the Dyson equation for the fermionic propagator is obtained as:
\be
G(\omega) = G^{0}(\omega) + G^{0}(\omega)\Sigma(\omega) G(\omega).
\label{Dyson}
\ee

The self-energy inherits the decomposition into the static and dynamical contributions from the $T$-matrix:
\be
\Sigma_{11'}(\omega) = \Sigma_{11'}^{0} + \Sigma_{11'}^{r}(\omega),
\label{Somega}
\ee
which can be specified after evaluating the commutators of Eqs. (\ref{Toperator}), namely
\be
 \Sigma^{0}_{11'} = -\langle[[V,\psi_1],{\psi^{\dagger}}_{1'}]_+\rangle = \sum\limits_{il}{\bar v}_{1i1'l}\rho_{li}, \ \ \ \ \ \rho_{li} = \langle{\psi^{\dagger}}_i\psi_l\rangle
 \label{MF}
\ee
with $\rho_{li}$ being the matrix elements of the ground-state one-body density, and $\Sigma_{11'}^{r}(\omega)$ as the Fourier transform of
\bea
\Sigma_{11'}^{r}(t-t') &=& -\frac{i}{4} \sum\limits_{npq}\sum\limits_{ikl}{\bar v}_{1ikl}\times \nonumber\\
&\times&\langle T \bigl(\psi^{\dagger}_i\psi_l\psi_k\bigr)(t)\bigl(\psi^{\dagger}_p\psi^{\dagger}_q\psi_n\bigr)(t')\rangle^{irr}
{\bar v}_{qpn1'}  \nonumber\\
&=& \frac{1}{4} \sum\limits_{npq}\sum\limits_{ikl}{\bar v}_{1ikl}
G^{(pph)irr}_{ilk,nqp}(t-t'){\bar v}_{qpn1'}.\nonumber\\
\label{Tr}
\eea
The static part $\Sigma^{0}$ (\ref{MF}), or the mean field, has the famous Hartree-Fock ansatz with, in general, correlated density, and the clearly separated dynamical part 
$\Sigma^{r}$ has, in this formulation, the symmetric form of a three-fermion CF "sandwiched" between two interaction matrix elements. Respectively, the former generates short-range correlations and the latter is responsible for the long-range ones. In this context, the short range is associated with the range of the bare interaction $\bar v$, and the long range extends to the size of the entire many-body system.
Eq. (\ref{Dyson}) is formally a closed exact equation for $G(\omega)$, however, it is complicated by the three-fermion CF in the dynamical kernel.

The analogous EOM method for the three-body two-time propagator generates a similar Dyson-type equation with static and dynamical self-energies. The latter is a function of even higher-rank propagators, which makes the exact solution of the many-body problem hardly tractable. Instead, approximations to the three-fermion propagator are considered. Within the non-perturbative paradigm, the cluster decomposition of the two-particle-one-hole CF in the dynamical kernel (\ref{Tr}) is one of the viable approaches. Symbolically, it takes the form:
\be
G^{(pph)irr} \sim G^{(p)}G^{(p)}G^{(h)} + G^{(p)}R^{(ph)} + G^{(h)}G^{(pp)} + \sigma^{(pph)},
\label{CD}
\ee
where the number of particles ($p$) and holes ($h$) in the superscripts indicates the rank of the respective CF, and the sum implicitly includes all the necessary antisymmetrizations. An accurate decomposition is given and discussed, e.g., in Refs. \cite{VinhMau1969,Mau1976,LitvinovaSchuck2019}. Truncation of the many-body problem at the one-body level confining by the first term on the right-hand side of Eq. (\ref{CD}), i.e., the self-consistent Green functions approach, is appropriate at weak coupling \cite{Poschenrieder1988,Poschenrieder1988a}. In this case, the single-fermion EOM acquires a closed form and admits iterative solutions. The next level of sophistication is decomposition, retaining all possible terms with one-fermion and two-fermion propagators, i.e., the second and third terms in Eq. (\ref{CD}). This class of solutions can be mapped to the PVC \cite{Martin1959,VinhMau1969,Mau1976,RingSchuck1980,Rijsdijk1992,LitvinovaSchuck2019}, including the coupling between particles and phonons of both particle-hole and particle-particle origins, and related to NFT \cite{BohrMottelson1969,BohrMottelson1975,Broglia1976,BortignonBrogliaBesEtAl1977,BertschBortignonBroglia1983}. 

The PVC concept will be the central one in the applications discussed in this course. Therefore, we dwell on its mathematical aspects and derive the explicit form of the PVC self-energy. 
After performing the decomposition (\ref{CD}) and dropping the last term (its role and quantitative contribution are commented on below), the one-fermion dynamical kernel takes the form
\begin{eqnarray}
  \Sigma^{r}_{11'}(\omega) =  \Sigma^{r(ph)}_{11'}(\omega) +  \Sigma^{r(pp)}_{11'}(\omega) +  \Sigma^{r(0)}_{11'}(\omega).\nonumber \\
 \label{SEirr2}
 \end{eqnarray}
The three terms on the right-hand side marked by '$(ph)$', '$(pp)$', and '$(0)$' are ordered according to their importance for typical nuclear structure calculations and correspond to the second, third, and first terms of Eq. (\ref{CD}), respectively. Explicitly, after performing the factorizations and Fourier transformations of the CF products to the energy domain, they read:
\bea
\Sigma^{r(ph)}_{11'}(\omega) = \sum\limits_{33'} \Bigl[ 
\sum\limits_{\nu n}\frac{\eta_3^{n}{g}_{13}^{\nu}{g}_{1'3'}^{\nu\ast}\eta_{3'}^{n\ast}}{\omega - \omega_{\nu} - \varepsilon_n^{(+)} + i\delta} +
\nonumber\\ +
\sum\limits_{\nu m} \frac{\chi_3^{m} g_{31}^{\nu\ast}g_{3'1'}^{\nu}\chi_{3'}^{m\ast}}{\omega + \omega_{\nu} + \varepsilon_m^{(-)} - i\delta} 
\Bigr],
\label{Sigmarph}
\eea
\bea
\Sigma^{r(pp)}_{11'}(\omega) = \sum\limits_{22'} \Bigl[ \sum\limits_{\mu m} \frac{\chi_2^{m\ast} \gamma_{12}^{\mu(+)}\gamma_{1'2'}^{\mu(+)\ast}\chi_{2'}^{m}}{\omega - \omega_{\mu}^{(++)} - \varepsilon_m^{(-)} + i\delta} + \nonumber\\
+ \sum\limits_{\varkappa n}\frac{\eta_2^{n\ast}{\gamma}_{21}^{\varkappa(-)\ast}{\gamma}_{2'1'}^{\varkappa(-)}\eta_{2'}^n}{\omega + \omega_{\varkappa}^{(--)} + \varepsilon_n^{(+)} - i\delta} \Bigr],
\label{FISrpp}
\eea
%
\bea
&\Sigma&^{r(0)}_{11'}(\omega) = -\sum\limits_{2342'3'4'} {\bar v}_{1234}\times \nonumber \\ &\times&\Bigl[\sum\limits_{mn'n''} \frac{\chi_{2'}^{m}\chi_2^{m\ast}\eta_3^{n'}\eta_{3'}^{n'\ast}\eta_4^{n''}\eta_{4'}^{n''\ast}}
{\omega - \varepsilon_{n'}^{(+)} - \varepsilon_{n''}^{(+)} - \varepsilon_{m}^{(-)} + i\delta} \nonumber \\
&+& \sum\limits_{nm'm''} \frac{\eta_{2'}^{n}\eta_{2}^{n\ast}\chi_3^{m'}\chi_{3'}^{m'\ast}\chi_4^{m''}\chi_{4'}^{m''\ast}}
{\omega + \varepsilon_{n}^{(+)} + \varepsilon_{m'}^{(-)} + \varepsilon_{m''}^{(-)} - i\delta} \Bigr] {\bar v}_{4'3'2'1'}, 
\label{Sigmar0}
\eea
where $\varepsilon_n^{(+)} = E^{(N+1)}_{n} - E^{(N)}_0$ are the single-particle energies in the neighboring $(N+1)$-particle system and 
$\varepsilon_m^{(-)} = E^{(N-1)}_{m} - E^{(N)}_0$ are those in the $(N-1)$-particle system. 

\begin{figure}
\begin{center}
\includegraphics[scale=0.52]{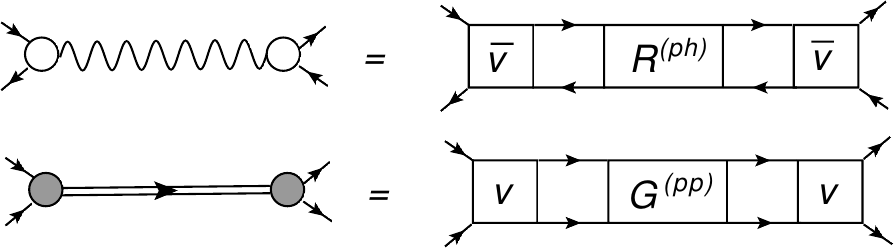}
\end{center}
\caption{The PVC "anatomy":  the normal (top) and superfluid (bottom) phonon vertices are denoted by the empty and filled circles, respectively, the wavy and double lines stand for their propagators. The bare interaction, antisymmetrized $\bar v$ and plain $v$,  is represented by the squares, and the two-fermion correlation functions are given by the rectangular blocks $R^{(ph)}$ and $G^{(pp)}$. Solid lines are associated with fermionic particle (right arrows) and hole (left arrows) states with respect to the Fermi energy. The figure is adapted from Ref. \cite{LitvinovaSchuck2019}.}
\label{PVCmap}%
\end{figure}

Mapping onto the PVC constitutes an important step that enables a reformulation of the many-body theory as an effective field theory. Unlike in most of the effective field theories, we can perform the mapping exactly by introducing the phonon degrees of freedom via the interaction amplitudes  
$\Gamma^{ph}$ and $\Gamma^{pp}$:
\bea
\Gamma^{ph}_{13',1'3} =  \sum\limits_{242'4'}{\bar v}_{1234}R^{(ph)}_{24,2'4'}(\omega){\bar v}_{4'3'2'1'} = \nonumber \\ = 
\sum\limits_{\nu,\sigma=\pm1} g^{{\nu}(\sigma)}_{13}D^{(\sigma)}_{\nu}(\omega)g^{\nu(\sigma)\ast}_{1'3'},
\label{mappingph}
\eea
\bea
\Gamma^{pp}_{12,1'2'}(\omega) = \sum\limits_{343'4'}{v}_{1234}G^{(pp)}_{43,3'4'}(\omega){v}_{4'3'2'1'} = \nonumber \\
= \sum\limits_{\mu,\sigma=\pm1} {\tilde\gamma}^{\mu(\sigma)}_{12}\Delta^{(\sigma)}_{\mu}(\omega){\tilde\gamma}^{\mu(\sigma)\ast}_{1'2'}.
\label{mappingpp}
\eea
The vertices of the normal (zero-sound) phonons $g^{\nu}$ and their propagators $D_{\nu}(\omega)$ are, thereby:
\bea
g^{\nu(\sigma)}_{13} = \delta_{\sigma,+1}g^{\nu}_{13} + \delta_{\sigma,-1}g^{\nu\ast}_{31}, \ \ \ \ 
g^{\nu}_{13} = \sum\limits_{24}{\bar v}_{1234}\rho^{\nu}_{42}, 
\label{vert_ph} \nonumber\\
\\
D_{\nu}^{(\sigma)}(\omega) = \frac{\sigma}{\omega - \sigma(\omega_{\nu} - i\delta)}, \ \ \ \
\omega_{\nu} = E_{\nu} - E_0, \nonumber \\
\label{gDPVCph}
\eea 
and
those for the superfluid pairing phonons $\gamma^{\mu(\pm)}$ and  $\Delta_{\mu}(\omega)$, respectively,  read:
\bea
{\tilde\gamma}^{\mu(\sigma)}_{12} = \gamma^{\mu(+)}_{12}\delta_{\sigma,1} +  \gamma^{\mu(-)\ast}_{12}\delta_{\sigma,-1}, \nonumber \\
\gamma^{\mu(+)}_{12} = \sum\limits_{34} v_{1234}\alpha_{34}^{\mu}, \ \ \ \ \ \ \gamma_{12}^{\varkappa(-)} = \sum\limits_{34}\beta_{34}^{\varkappa}v_{3412}, 
\label{vert_pp}
\eea
\be
\Delta^{(\sigma)}_{\mu}(\omega) = \frac{\sigma}{\omega - \sigma(\omega_{\mu}^{(\sigma\sigma)} - i\delta)}.
\ee
In the expressions above, the particle-particle ($pp$) and particle-hole ($ph$) correlation functions of Eqs. (\ref{phresp}, \ref{ppGF}) are employed. The PVC mappings introduced by Eqs. (\ref{mappingph}, \ref{mappingpp}) are displayed diagrammatically in Fig. \ref{PVCmap}.
%

\begin{figure*}
\begin{center}
\includegraphics*[scale=0.75]{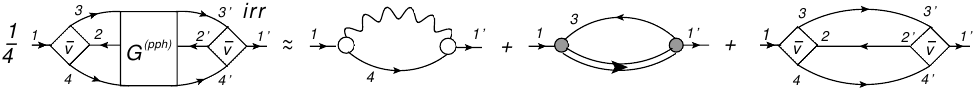}
\end{center}
\caption{The dynamical self-energy ( kernel $\Sigma^{r}$ of Eq. (\ref{SEirr2}) in terms of PVC, using the same conventions as in Fig. \ref{PVCmap}. The block $G^{(pph)}$ represents the three-fermion propagator of Eq. (\ref{Tr}). The figure is adapted from Ref. \cite{LitvinovaSchuck2019}.}
\label{SEdyn}%
\end{figure*}

The mappings (\ref{mappingph}, \ref{mappingpp}) can further link the spectral expansions (\ref{Sigmarph} -- \ref{Sigmar0}) to the diagrammatic form of the dynamical self-energy shown in Fig. \ref{SEdyn}. Here we notice that its three terms correspond to two one-loop and one two-loop diagrams which may be linked to the Feynman diagrams and, thus, evaluated by the Feynman rules:
\be
\Sigma^{r(ph)}_{11'}(\omega) = -\sum\limits_{33'}\int\limits_{-\infty}^{\infty}\frac{d\varepsilon}{2\pi i} \Gamma^{ph}_{13',1'3}(\omega - \varepsilon)G_{33'}(\varepsilon),
\label{FISphe}
\ee
\be
\Sigma^{r(pp)}_{11'}(\omega) = \sum\limits_{22'}\int\limits_{-\infty}^{\infty}\frac{d\varepsilon}{2\pi i} \Gamma^{pp}_{12,1'2'}(\omega + \varepsilon)G_{2'2}(\varepsilon),
\ee
\bea
\Sigma^{r(0)}_{11'}(\omega) &=& -\sum\limits_{2342'3'4'}{\bar v}_{1234}\nonumber\\
&\times~&\int\limits_{-\infty}^{\infty}\frac{d\varepsilon d\varepsilon'}{(2\pi i)^2}  G_{44'}(\omega+\varepsilon'-\varepsilon)G_{33'}(\varepsilon)G_{2'2}(\varepsilon')
\nonumber \\
&\times&{\bar v}_{4'3'2'1'}.
\label{SEdyn0}
\eea

Here, we make a remark that the signs in front of the diagrams may be convention-dependent. The last term, in particular, may be given with the "-" sign in some literature because this sign shows up explicitly in Eq. (\ref{Sigmar0}). Furthermore, the phonon vertices may include phase factors. 
In Fig. \ref{SEdyn}, the first two diagrams on the right-hand side have the topology analogous to the electron self-energy lowest-order correction in quantum electrodynamics. The emission and reabsorption of a lattice phonon excitation is the process corresponding to such one-loop self-energies in a solid, and meson exchange is a matching case from quantum hadrodynamics. In the nuclear structure applications, 
the one-meson exchange acts as a bare interaction (or that with slightly renormalized coupling constants), while multiple in-medium meson-exchange processes are generated by the EOM non-perturbatively. 
In a strong coupling regime, the leading approximation is dominated by the term $\Sigma^{r(ph)}$ (\ref{Sigmarph}), which takes the form analogous to the boson exchange, thereby illustrating how the fermionic interaction is driven across the energy scales. The possibility of the PVC mapping signals the appearance of the new order parameter associated with the coupling vertex, which {\it takes over the power counting governing the bare nucleon-nucleon interaction evaluated with the vacuum averages}. In the many-body system, the latter is replaced by the ground-state averages, which provide an adequate framework where the description can be organized around the mean field emerging from collective effects.

After the PVC mapping, an effective Hamiltonian with explicit phonon degrees of freedom can be introduced, leading to NFT, the theory of nucleons and phonons in a correlated medium. In contrast to QFTs, where the coupled fermionic and bosonic degrees of freedom are independent, in NFT, the phonons are composed of correlated fermionic pairs. Furthermore, as we have seen, the PVC vertices in nuclear systems are not the effective parameters of the theory but are, in principle, calculable from the underlying fermionic bare interaction, i.e., NFT can be "ultraviolet (UV) complete" up to the underlying bare NN interaction defined for nucleons in the vacuum. Of course, in our case, all the expectation values are formally taken in correlated many-body states, while QFTs operate vacuum averages.


Some realizations of the PVC model based on the bare NN interaction are available in Refs. \cite{Barbieri2009,Barbieri2009a}. Quantitatively, implementations that use effective interactions \cite{LitvinovaRing2006,Litvinova2012,LitvinovaAfanasjev2011,AfanasjevLitvinova2015} are more accurate in comparison to data. They are also easier to implement as the phonon characteristics can be found on the one-loop level of (quasiparticle) RPA ((Q)RPA), therefore, an arduous task of solving a non-linear  Dyson equation for the single-particle GF can be avoided.  The use of an effective interaction in conjunction with PVC requires a refinement to remove the double counting of the PVC effects, implicitly contained in the static kernel. The most efficient method is subtracting the static, or zero frequency, limit of the dynamical PVC contribution from the total kernel \cite{Tselyaev2013} which was shown to be well behaving in the computation of the particle-hole response \cite{Tselyaev2013,LitvinovaTselyaev2007,LitvinovaRingTselyaev2007,LitvinovaRingTselyaev2008, Gambacurta2015,LitvinovaSchuck2019}. Furthermore, this method prevents the shift of the (Q)RPA Goldstone modes with PVC and improves the convergence of the two-fermion dynamical kernels in self-consistent computational schemes. Practical aspects of the subtraction are discussed in Section \ref{calculations}.

Resuming the discussion of the complete dynamical kernel (\ref{SEirr2}), in theory, truncated on the two-body level, it requires the external input in terms of $R^{(ph)}$ and $G^{(pp)}$ CFs and the extraction of phonon vertices and frequencies.  To obtain these characteristics the corresponding EOMs for these CFs should be solved separately. The detailed formalism for the two-body sector of the theory and some results can be found in Refs. \cite{DukelskyRoepkeSchuck1998,Olevano2018,LitvinovaSchuck2019,LitvinovaSchuck2020}. The response theory is the subject of Section \ref{response}, and its implementations are discussed in Section \ref{calculations}.


The different sign of the "second-order" term $\Sigma^{r(0)}$ containing only the single-fermion GFs, with respect to the "radiative correction" terms $\Sigma^{r(ph)}$ and $\Sigma^{r(pp)}$ involving contractions with correlated fermionic pairs, indicates that the positivity condition can be, in principle, violated, which causes problems with the optical theorem \cite{Schuck1973,Danielewicz1994}. In approximations of the intermediate and strong coupling, however, the $\Sigma^{r(0)}$ term is generally smaller than the other two encompassing collective effects. In other settings, preserving the integrity of the spectral decomposition of $\Sigma^{r}$ guarantees the positivity as
\bea
\Sigma_{11'}^{r}(\omega) &=& \frac{1}{4} \sum\limits_{rpq}\sum\limits_{ikl}{\bar v}_{1ikl}
G^{(pph)irr}_{ilk,rqp}(\omega){\bar v}_{qpr1'}\nonumber\\
\label{Tromega}
G^{(pph)}_{ilk,rqp} (\omega) &=&  \sum\limits_{n}\frac{\langle 0|\psi^{\dagger}_i\psi_{l}\psi_{k}|n\rangle\langle n|\psi^{\dagger}_{p}\psi^{\dagger}_{q}\psi_{r}|0\rangle}{\omega - (E^{(N+1)}_{n} - E^{(N)}_0)+i\delta} \nonumber \\
&+& \sum\limits_{m}\frac{\langle 0| \psi^{\dagger}_{p}\psi^{\dagger}_{q}\psi_{r}|m\rangle\langle m|\psi^{\dagger}_i\psi_{l}\psi_{k} |0\rangle}{\omega + (E^{(N-1)}_{m} - E^{(N)}_0)-i\delta}. \nonumber \\ 
\label{pphgfspec}
\eea
Some attempts to build approximate solutions for the irreducible part of $G^{(pph)}$ respecting the algebraic structure of the spectral series (\ref{pphgfspec}) are known from the literature. The two-particle-one-hole ($2p1h$) RPA for this CF  was elaborated in Ref. \cite{Schuck1973}, and Faddeev series were considered for the energies and matrix elements of the correlated $2p1h$ configurations in Ref. \cite{Danielewicz1994}. The importance of the $ph$ and $pp$ emergent collective effects in the dynamical self-energy was particularly recognized.
Some numerical studies within the Tamm-Dancoff approach and RPA for $ph$ and $pp$ CFs in nucleonic self-energies can be found in Refs.\cite{Rijsdijk1992,Rijsdijk1996,Barbieri2009,Barbieri2009a}. 

\subsection{Single-fermion EOM in the superfluid phase}
\label{superfluid}

Superfluidity is a pronounced emergent property of strongly correlated fermionic systems, including atomic nuclei \cite{RingSchuck1980}. The inclusion of superfluidity requires an extended treatment on the theory side. The superfluid phase of fermionic systems exhibits an enhanced formation of Cooper pairs and their dynamical counterparts pairing phonons. The latter already appeared in the $\Sigma^{r(pp)}$ part of the fermionic dynamical self-energy derived in the previous subsection. When an effective interaction is used to describe the static part of the NN in-medium self-energy, the quantitative contribution of the pairing term is smaller than that of 
$\Sigma^{r(ph)}$ in the nuclear systems with double shell closure where the formation of the pairing gaps is suppressed by large shell gaps. It is, however, enhanced in open-shell systems, and the relative importance of  $\Sigma^{r(pp)}$ and $\Sigma^{r(ph)}$ may change in computation with a bare interaction. Moreover, these two terms form a multi-component structure that can be unified in the formalism of Bogolyubov's quasiparticles. 
 In the PVC approach discussed above, the superfluid pairing is fully dynamic as it is mediated by the pairing phonons emerging in the one-fermion self-energy from the bare interaction. In a framework based on effective interaction, the superfluid pairing can be included statically in the Bardeen-Cooper-Schrieffer (BCS) approximation or the Hartree-(Fock)-Bogolyubov one. The corresponding Green function technique is the Gor'kov Green function, which extends the notion of the one-fermion propagator (\ref{spgf}) by introducing anomalous components with the same kind of fermionic operators, which is done below. These CFs are non-negligible because of the presence of correlated fermionic pairs in the ground state of the system. We will also see that more two-body CFs should be retained in the dynamical self-energy.

The simplest approach to the single-fermion motion with superfluidity can be obtained from the EOM1 neglecting two-body and higher-rank CFs \cite{Gorkov1958}. In Ref. \cite{Litvinova2021a}, Gor'kov theory is generalized beyond this approximation to the inclusion of the PVC effects. Here we briefly discuss this approximation, for which
it is convenient to introduce the HFB basis, or the basis of the Bogolyubov quasiparticles \cite{Bogolubov1947}. The single-fermion states in this basis are superpositions of
particles and holes, i.e., the fermionic states above and below the Fermi energy: 
\bea
\psi_1 = \sum\limits_{\mu} \bigl(U_{1\mu}\alpha_{\mu} + V^{\ast}_{1\mu}\alpha^{\dagger}_{\mu}\bigr) \nonumber\\
\psi^{\dagger}_1 = \sum\limits_{\mu} \bigl(V_{1\mu}\alpha_{\mu} + U^{\ast}_{1\mu}\alpha^{\dagger}_{\mu}\bigr),
\label{Btrans}
\eea
where $\alpha$ and $\alpha^{\dagger}$ are the quasiparticle operators obeying the same anticommutator algebra as the particle operators $\psi$ and $\psi^{\dagger}$.
Here and henceforth, the fermionic states in the HFB basis are marked by the Greek indices while the single-particle (mean-field) basis states are associated with the number and Roman indices. Eq. (\ref{Btrans}) is known in an array form:
\bea
\left( \begin{array}{c} \psi \\ \psi^{\dagger} \end{array} \right) = \cal{W} \left( \begin{array}{c} \alpha \\ \alpha^{\dagger} \end{array} \right),
\eea
with the unitary matrices
\bea
\cal{W} = \left( \begin{array}{cc} U & V^{\ast} \\ V & U^{\ast} \end{array} \right) \ \ \ \ \ \  \cal{W}^{\dagger} = \left( \begin{array}{cc} U^{\dagger} & V^{\dagger} \\ V^T & U^T \end{array} \right). 
\eea
The matrix blocks $U$ and $V$ satisfy the following conditions \cite{RingSchuck1980}:
\bea
U^{\dagger}U + V^{\dagger}V = \mathbb{1}\ \ \ \ \ \ UU^{\dagger} + V^{\ast}V^{T} = \mathbb{1}\nonumber\\
U^TV + V^TU = 0\ \ \ \ \ \  UV^{\dagger} + V^{\ast}U^{T} = 0 .
\label{UV}
\eea
It is convenient to consider a generalized four-component, or {\it quasiparticle}, fermionic propagator
\bea
{\hat G}_{12}(t-t') = -i\langle T\Psi_1(t)\Psi^{\dagger}_2(t')\rangle = \nonumber\\
= -i\theta(t-t')\left( \begin{array}{cc} \langle \psi_1(t)\psi^{\dagger}_2(t')\rangle &  \langle \psi_1(t)\psi_2(t')\rangle \\
 \langle \psi^{\dagger}_1(t)\psi^{\dagger}_2(t')\rangle &  \langle \psi^{\dagger}_1(t)\psi_2(t')\rangle
\end{array} \right)  + \nonumber \\
+ i\theta(t'-t)\left( \begin{array}{cc} \langle \psi^{\dagger}_2(t')\psi_1(t)\rangle &  \langle \psi_2(t')\psi_1(t)\rangle \\
 \langle \psi^{\dagger}_2(t')\psi^{\dagger}_1(t)\rangle &  \langle \psi_2(t')\psi^{\dagger}_1(t)\rangle
\end{array} \right) \nonumber\\
= \left( \begin{array}{cc} G_{12}(t-t')  &  F^{(1)}_{12}(t-t') \\ 
F^{(2)}_{12}(t-t') & G^{(h)}_{12}(t-t')\end{array}\right), 
\label{GG}
\eea
via introducing
\be
\Psi_1(t) = \left( \begin{array}{c} \psi_1(t) \\ \psi_1^{\dagger}(t) \end{array} \right), \ \ \ \ \ \ \ \ \ 
\Psi^{\dagger}_1(t) = \Bigl( \psi_1^{\dagger}(t) \ \ \ \ \ \psi_1(t) \Bigr).
\label{psicol}
\ee
For each component of the propagator (\ref{GG}), one can generate an EOM in a similar fashion as in the previous subsection. 
The resulting 2$\times$2 matrix equation for the ${\hat G}_{12}$ can be subsequently transformed to the quasiparticle basis  \cite{Litvinova2021a}.
This yields a formally more compact Gor'kov-Dyson equation for the forward $\eta = +$ and backward $\eta = -$ components of the quasiparticle propagator:
\be
G^{(\eta)}_{\nu\nu'}(\varepsilon) = {\tilde G}^{(\eta)}_{\nu\nu'}(\varepsilon) + \sum\limits_{\mu\mu'}{\tilde G}^{(\eta)}_{\nu\mu}(\varepsilon)\Sigma^{r(\eta)}_{\mu\mu'}(\varepsilon)G^{(\eta)}_{\mu'\nu'}(\varepsilon). 
\label{Dyson_qp}
\ee
Its mean-field, or uncorrelated, part ${\tilde G}^{(\eta)}_{\nu\nu'}(\varepsilon)$ absorbs the static kernel and can be made diagonal by a proper choice of the working basis:
\be
{\tilde G}^{(\eta)}_{\nu\nu'}(\varepsilon) = \frac{\delta_{\nu\nu'}}{\varepsilon - \eta(E_{\nu} - E_0 - i\delta)},
\label{MFG_qp}
\ee
while generally non-diagonal correlated  propagator $G^{(\eta)}_{\nu\nu'}(\varepsilon)$ reads:
\be
{G}^{(\eta)}_{\nu\nu'}(\varepsilon) = \sum\limits_n\frac{S^{\eta(n)}_{\nu\nu'}}{\varepsilon - \eta(E_{n} - E_0 - i\delta)}.
\label{G_qp}
\ee
In Eqs. (\ref{MFG_qp}, \ref{G_qp}), $E_{\nu}$ are the mean-field energies of the Bogolyubov quasiparticles and $E_{n}$ are their energies affected by beyond-mean-field correlations.
The residues ${S^{\eta(n)}_{\nu\nu'}}$ are the basis-dependent spectroscopic factors: $S^{+(n)}_{\nu\nu'} = \langle 0|\alpha_{\nu}|n\rangle\langle n|\alpha^{\dagger}_{\nu'}|0\rangle$ and $S^{-(m)}_{\nu\nu'} = \langle 0|\alpha_{\nu}|m\rangle\langle m|\alpha^{\dagger}_{\nu'}|0\rangle$ with the correlated (formally exact) many-body states $|n\rangle$ and $|m\rangle$. 
The two representations of the dynamical kernel are related as follows:
\bea
\Sigma^{r(+)}_{\mu\mu'}(\varepsilon) &=& \sum\limits_{12} \Bigl(U^{\dagger}_{\mu 1} \ \ \  V^{\dagger}_{\mu 1} \Bigr) \left( \begin{array}{cc} \Sigma^r_{12}(\varepsilon) &  \Sigma^{(1)r}_{12}(\varepsilon)\\ \Sigma^{(2)r}_{12}(\varepsilon) & \Sigma^{(h)r}_{12}(\varepsilon)\end{array}\right)
\left( \begin{array}{c} U_{2\mu'} \\ V_{2\mu'} \end{array}\right)  \nonumber\\
&=& \sum\limits_{12} \Bigl( U^{\dagger}_{\mu 1}\Sigma^r_{12}(\varepsilon)U_{2\mu'} + U^{\dagger}_{\mu 1}\Sigma^{(1)r}_{12}(\varepsilon)V_{2\mu'} \nonumber\\
&+& V^{\dagger}_{\mu 1}\Sigma^{(2)r}_{12}(\varepsilon)U_{2\mu'} + V^{\dagger}_{\mu 1}\Sigma^{(h)r}_{12}(\varepsilon)V_{2\mu'} \Bigr),
\label{Sigma+}
\eea
\bea
\Sigma^{r(-)}_{\mu\mu'}(\varepsilon) &=& \sum\limits_{12} \Bigl(V^{T}_{\mu 1} \ \ \  U^{T}_{\mu 1} \Bigr) \left( \begin{array}{cc} \Sigma^r_{12}(\varepsilon) &  \Sigma^{(1)r}_{12}(\varepsilon)\\ \Sigma^{(2)r}_{12}(\varepsilon) & \Sigma^{(h)r}_{12}(\varepsilon)\end{array}\right)
\left( \begin{array}{c} V^{\ast}_{2\mu'} \\ U^{\ast}_{2\mu'} \end{array}\right)  \nonumber\\
&=& \sum\limits_{12} \Bigl( V^{T}_{\mu 1}\Sigma^r_{12}(\varepsilon)V^{\ast}_{2\mu'} + V^{T}_{\mu 1}\Sigma^{(1)r}_{12}(\varepsilon)U^{\ast}_{2\mu'} \nonumber\\
&+& U^{T}_{\mu 1}\Sigma^{(2)r}_{12}(\varepsilon)V^{\ast}_{2\mu'} + U^{T}_{\mu 1}\Sigma^{(h)r}_{12}(\varepsilon)U^{\ast}_{2\mu'} \Bigr),
\label{Sigma-}
\eea
with the obvious correspondence between the block matrix structure of the dynamical self-energy and that of the propagator matrix (\ref{GG}). 
%
\begin{figure*}
\begin{center}
\vspace{0.3cm}
\includegraphics[scale=0.8]{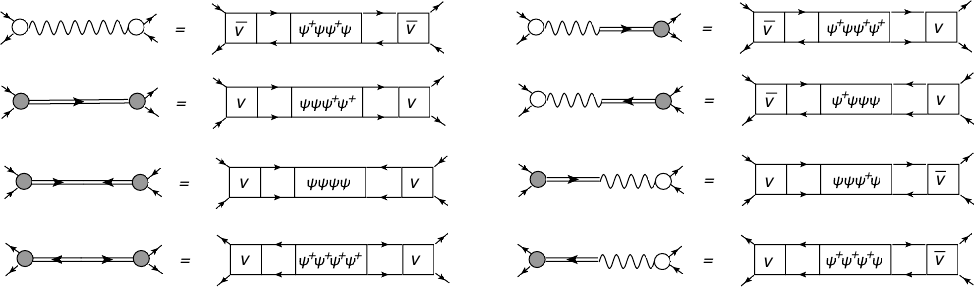}
\end{center}
\caption{The complete set of the qPVC amplitudes in the diagrammatic representation with the same conventions as in Figs. \ref{PVCmap}, \ref{SEdyn}. 
The depicted two-body two-point correlation functions are associated with the operator products given in the rectangular boxes, according to the rule: \framebox{abcd} $= -i\langle T(ab)(t)(cd)(t')\rangle$, and include the attached fermionic lines (arrowed lines). The figure is adapted from Ref.  \cite{Zhang2022}.}
\label{QVC_map}
\end{figure*}

The double contractions of the three-fermion CFs with two interaction matrix elements give the four time-dependent components of the exact dynamical self-energy in the single-particle basis:
\bea
\Sigma_{11'}^{r}(t-t') &=& \frac{i}{4} \sum\limits_{ikl}\sum\limits_{mnq}{\bar v}_{1ikl} \nonumber\\
&\times&\langle T \bigl(\psi^{\dagger}_i\psi_l\psi_k\bigr)(t)\bigl(\psi^{\dagger}_m\psi^{\dagger}_n\psi_q\bigr)(t')\rangle^{irr}
{\bar v}_{mnq1'} \nonumber\\
\Sigma_{11'}^{(1)r}(t-t') 
&=& \frac{i}{4}\sum\limits_{ikl}\sum\limits_{mnq}{\bar v}_{1ikl} \nonumber\\
&\times&\langle T(\psi^{\dagger}_i\psi_l\psi_k )(t)(\psi^{\dagger}_m\psi_q\psi_n)(t')\rangle^{irr} {\bar v}_{1'mnq}. \nonumber \\
\Sigma_{11'}^{(2)r}(t-t') 
&=& \frac{i}{4}\sum\limits_{ikl}\sum\limits_{mnq}{\bar v}_{ikl1} \nonumber\\
&\times&\langle T(\psi^{\dagger}_i\psi^{\dagger}_k\psi_l )(t)(\psi^{\dagger}_m\psi^{\dagger}_n\psi_q)(t')\rangle^{irr} {\bar v}_{mnq1'}\nonumber \\
\Sigma_{11'}^{(h)r}(t-t') 
&=& \frac{i}{4}\sum\limits_{ikl}\sum\limits_{mnq}{\bar v}_{ikl1} \nonumber\\
&\times&\langle T(\psi^{\dagger}_i\psi^{\dagger}_k\psi_l )(t)(\psi^{\dagger}_m\psi_q\psi_n)(t')\rangle^{irr} {\bar v}_{1'mnq}. \nonumber \\
\label{Tsfdyn}
\nonumber\\
\eea
%
\begin{figure*}
\begin{center} 
\includegraphics[scale=0.34]{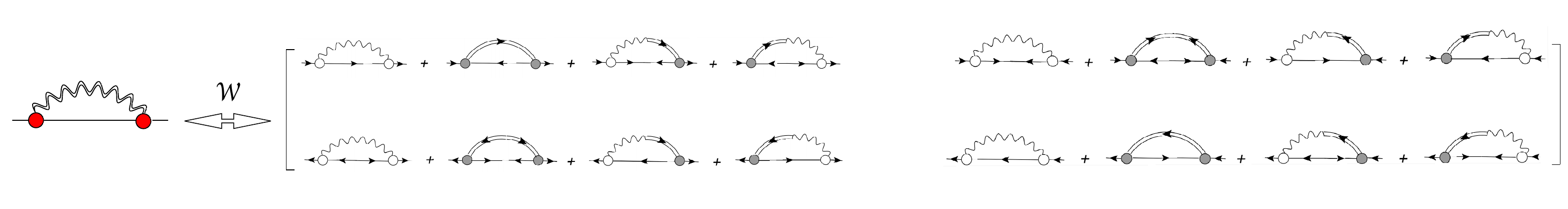}
\vspace{-1.0cm}
\end{center}
\caption{The superfluid dynamical self-energy in the quasiparticle (left) and single-particle (right) bases, related by Bogolyubov's transformation $\cal W$, in the qPVC approach. The double wavy line stands for the propagator of a unified superfluid phonon in the quasiparticle basis, the filled (red) circles are reserved for the respective combined qPVC phonon vertices, and the single line without arrows is attributed to the quasiparticle propagator unifying particles and holes. The remaining conventions are kept intact with the preceding figures. The figure is adapted from Ref. \cite{Litvinova2022}.}
\label{SE_qvc}%
\end{figure*}

Similarly to the normal case, the cluster decomposition (\ref{CD}) can be applied to the CFs of Eq. (\ref{Tsfdyn}). At this point, one realizes that more two-fermion CFs give non-vanishing contributions in the superfluid ground state. Their contributions are listed in Fig. \ref{QVC_map} in the diagrammatic form.
Accordingly, more of the corresponding EOMs should be supplied to the dynamical kernel, which may look overwhelming. However, rescue comes with transforming the Dyson equation to the quasiparticle basis (\ref{Dyson_qp} - \ref{Sigma-}). As a result of this transformation,  the dynamical kernel $\Sigma^{r(+)}_{\nu\nu'}(\varepsilon)$ takes the form 
\bea
\Sigma^{r(+)}_{\nu\nu'}(\varepsilon) = \sum\limits_{\nu''\mu} \Bigl[ 
\frac{{\Gamma}^{(11)\mu}_{\nu\nu''}{\Gamma}^{(11)\mu\ast}_{\nu'\nu''}}{\varepsilon  - E_{\nu''} - \omega_{\mu} + i\delta} +
\frac{{\Gamma}^{(02)\mu\ast}_{\nu\nu''}{\Gamma}^{(02)\mu}_{\nu'\nu''}}{\varepsilon + E_{\nu''} + \omega_{\mu} - i\delta} \Bigl], \nonumber\\
\label{SEqp}
\eea
where the entire multi-component structure of the two-fermion GFs listed on the right-hand side of Fig. \ref{QVC_map} becomes packed into the vertex functions $\Gamma^{(11)}$ and $\Gamma^{(02)}$ defined as follows:
\bea
\Gamma^{(11)\mu}_{\nu\nu'} = \sum\limits_{12}\Bigl[ 
U^{\dagger}_{\nu 1}(g^{\mu}_{12}\eta^{\nu'}_2 + \gamma^{\mu(+)}_{12}\chi^{\nu'\ast}_2) - \nonumber \\
- V^{\dagger}_{\nu 1}((g^{\mu }_{12})^T\chi^{\nu'\ast}_2 + (\gamma^{\mu(-)}_{12})^T\eta^{\nu'}_2)\Bigr]
\label{Gamma11_qp}
\\
\Gamma^{(02)\mu}_{\nu\nu'} = -\sum\limits_{12}\Bigl[ 
V^{T}_{\nu 1}(g^{\mu}_{12}\eta^{\nu'}_2 + \gamma^{\mu(+)}_{12}\chi^{\nu'\ast}_2) - \nonumber \\
- U^{T}_{\nu 1}((g^{\mu }_{12})^T\chi^{\nu'\ast}_2 + (\gamma^{\mu(-)}_{12})^T\eta^{\nu'}_2)\Bigr].
\label{Gamma02_qp}
\eea
The $\eta = -$ counterpart is an analog of Eq. (\ref{SEqp}) and yields corresponding opposite-sign energy solutions, which makes it practically redundant. Further, the simplest HFB approximation to the matrix elements $\eta^{\nu}_{i}$ and  $\chi^{\nu}_{i}$ reduces Eqs. (\ref{Gamma11_qp}) and  (\ref{Gamma02_qp}) to
\bea
\Gamma^{(11)\mu}_{\nu\nu'} = 
\Bigl[ 
U^{\dagger}g^{\mu}U + U^{\dagger}\gamma^{\mu(+)}V  \nonumber\\
- V^{\dagger}g^{\mu T}V - V^{\dagger}\gamma^{\mu(-)T}U\Bigr]_{\nu\nu'} \nonumber\\
\label{Gamma11_HFB}
\eea
\bea
\Gamma^{(02)\mu}_{\nu\nu'} =-\Bigl[ 
V^{T}g^{\mu}U+ V^{T}\gamma^{\mu(+)}V \nonumber\\
- U^{T}g^{\mu T}V - U^{T}\gamma^{\mu(-)T}U\Bigr]_{\nu\nu'}.\nonumber\\
\label{Gamma02_HFB}
\eea
Summarizing, after the factorization retaining all relevant two-fermion CFs, we arrive at the Gor'kov-Dyson equation taking a compact form in the HFB basis (\ref{Dyson_qp}). In this case, the dynamical kernel (\ref{SEqp}) has the same algebraic ansatz as in the non-superfluid case, and the additional complexity of the theory is transferred to the structure of the qPVC vertices  (\ref{Gamma11_qp}, \ref{Gamma02_qp}). The corresponding operation is shown in Fig. \ref{SE_qvc}. Remarkably, in this formulation, the superfluid phonons represent a unification of the normal and pairing phonons, whose vertices were shown to be proportional to the variations of the HFB Hamiltonian  \cite{Litvinova2021a}. Solutions of Eq. (\ref{Dyson_qp}) enable extraction of the single-particle characteristics of the many-body system. Refs. \cite{Litvinova2012,AfanasjevLitvinova2015} discuss the numerical solutions for spherical heavy nuclei, which allow for the shell evolution analyses and understanding of some features of spin-orbit splitting based on the "non-linear sigma" effective meson-exchange interactions adjusted within the covariant DFT (NL3 and NL3$^{\ast}$) \cite{Lalazissis1997,NL3star}.  Ref. \cite{Zhang2022} reports on an analogous implementation for axially deformed nuclei using the density-dependent meson exchange \cite{VretenarAfanasjevLalazissisEtAl2005}. In both cases, the density of single-quasiparticle states is too small in the relativistic H(F)B mean-field approximation, and the inclusion of the dynamical kernels in the leading-order qPVC improves the description considerably.   

\subsection{The superfluid response theory}
\label{response}

In the qPVC approach, the presence of the two-fermion propagators (\ref{phresp}, \ref{ppGF}) in the dynamical kernel $\Sigma^r$ of the single-particle EOM (\ref{Dyson}, \ref{Dyson_qp}) requires the input in terms of these CFs, which is external to the single-particle EOM. In the explicit expressions for $\Sigma^r$ (\ref{Sigmarph}, \ref{FISrpp}, \ref{SEqp}), one finds qPVC vertices and phonon frequencies. The latter are the poles of the $ph$ and $pp$ ($2q$) propagators, and the former are obtained by contractions of the CFs residues with the bare fermionic interaction.  The most direct way to obtain these elements of the qPVC amplitudes is to solve the EOMs for the CFs (\ref{phresp}, \ref{ppGF}).
These EOMs were formulated, e.g., in Refs. \cite{AdachiSchuck1989,DukelskyRoepkeSchuck1998}, and their dynamical kernels in the PVC approximation were derived, implemented, and discussed in Refs. \cite{LitvinovaSchuck2019,Schuck2019,LitvinovaSchuck2020}.
Similar to the one-body case, the two-body propagators of Eqs. (\ref{phresp}, \ref{ppGF}) are expected to undergo unification in the superfluid theory, however, this task becomes quite cumbersome in terms of the multicomponent Gor'kov Green functions (\ref{GG}). The previous subsection suggests that the formalism can become simpler in a quasiparticle basis. Therefore, we derive the EOMs for the two-fermion CFs in the quasiparticle space introduced by Eq. (\ref{Btrans}) from the beginning. The detailed derivation can be found in Ref. \cite{Litvinova2022}, and here I walk the reader through the major steps and outline feasible approximations of the qPVC class. 

The Hamiltonian (\ref{Hamiltonian}) acquires the following form in terms of the quasiparticle operators \cite{RingSchuck1980}:
\bea
H = H^0 &+& \sum\limits_{\mu\nu}H^{11}_{\mu\nu}\alpha^{\dagger}_{\mu}\alpha_{\nu}  + \frac{1}{2}\sum\limits_{\mu\nu}\bigl(H^{20}_{\mu\nu}\alpha^{\dagger}_{\mu}\alpha^{\dagger}_{\nu} + \text{h.c.}\bigr) \nonumber\\
&+& \sum\limits_{\mu\mu'\nu\nu'}\bigl(H^{40}_{\mu\mu'\nu\nu'}\alpha^{\dagger}_{\mu}\alpha^{\dagger}_{\mu'}\alpha^{\dagger}_{\nu}\alpha^{\dagger}_{\nu'} + \text{h.c}\bigr) \nonumber\\
&+& \sum\limits_{\mu\mu'\nu\nu'}\bigl(H^{31}_{\mu\mu'\nu\nu'}\alpha^{\dagger}_{\mu}\alpha^{\dagger}_{\mu'}\alpha^{\dagger}_{\nu}\alpha_{\nu'} + \text{h.c}\bigr) 
\nonumber\\
&+& \frac{1}{4}\sum\limits_{\mu\mu'\nu\nu'}H^{22}_{\mu\mu'\nu\nu'}\alpha^{\dagger}_{\mu}\alpha^{\dagger}_{\mu'}\alpha_{\nu'}\alpha_{\nu}, 
\label{Hqua}
\eea
where the upper indices in the Hamiltonian matrix elements $H^{ij}_{\mu\nu}$ and  $H^{ij}_{\mu\nu\mu'\nu'}$ correspond to the numbers of creation and annihilation quasiparticle operators in the associated terms. The matrix elements $H^{ij}$ in terms of the bare interaction and $\cal W$ transformation coefficients can be found in \cite{RingSchuck1980}. The quantity $H^{20}$ vanishes at the stationary point of the HFB equations, and the matrix elements of $H^{11}$ are attributed to the quasiparticle energies, that is $H^{11}_{\mu\nu} = \delta_{\mu\nu}E_{\mu}$:
 \be
 H = H^0 + \sum\limits_{\mu}E_{\mu}\alpha^{\dagger}_{\mu}\alpha_{\mu} + V,
 \label{Hqua1}
 \ee
where the residual interaction $V$ includes the remaining terms (the last three lines of Eq. (\ref{Hqua})). 

Now we depart from considering the closed system and subject it to an external field $F$, which is supposed to be sufficiently weak. 
Under this condition, the transitions of the many-body system to excited states can be treated in the lowest order perturbation theory (with respect to the field $F$). The strength function
defined as
\be
S(\omega) = \sum\limits_{n>0} \Bigl[ |\langle n|F^{\dagger}|0\rangle |^2\delta(\omega-\omega_n) - |\langle n|F|0\rangle |^2\delta(\omega+\omega_n)
\Bigr],
\label{SF}
\ee
where the summation runs over all the formally exact excited states $|n\rangle$, describes the spectral probability distribution. 

The one-body operator $F$ is canonically expanded in terms of the quasiparticle field operators:
\bea
F = \frac{1}{2}\sum\limits_{\mu\mu'} \Bigl(F^{20}_{\mu\mu'}\alpha^{\dagger}_{\mu}\alpha^{\dagger}_{\mu'} + 
F^{02}_{\mu\mu'}\alpha_{\mu'}\alpha_{\mu} \Bigr)\nonumber\\
F^{\dagger} = \frac{1}{2}\sum\limits_{\mu\mu'} \Bigl(F^{20\ast}_{\mu\mu'}\alpha_{\mu'}\alpha_{\mu} +
F^{02\ast}_{\mu\mu'}\alpha^{\dagger}_{\mu}\alpha^{\dagger}_{\mu'}  
\Bigr),
\label{Fext}
\eea
which is obtained via the Bogolyubov's transformation of the second-quantized form of $F = \sum_{12}F_{12}\psi^{\dagger}_1\psi_2$ after dropping the $F^{00}$ term not generating excited states and $F^{11}$ not contributing in the linear approximation  \cite{Avogadro2011}. Subleading contributions will be discussed elsewhere. Eq. (\ref{SF}) takes the form:
\bea
S(\omega) &=& -\frac{1}{\pi}\lim\limits_{\Delta \to 0} \text{Im} \Pi(\omega),\nonumber\\
\label{SFDelta} 
\Pi(\omega) &=&  \frac{1}{4}\sum\limits_{\mu\mu'\nu\nu'}
\left(\begin{array}{cc} F^{02}_{\mu\mu'} & F^{20}_{\mu\mu'} \end{array}\right){\hat{\cal R}}_{\mu\mu'\nu\nu'}(\omega+i\Delta)\left(\begin{array}{c} F^{02\ast}_{\nu\nu'} \\  \\ F^{20\ast}_{\nu\nu'} \end{array}\right),\nonumber\\
\label{Polar}
\eea
which sets the ansatz for the response function block matrix:
\bea
{\hat{\cal R}}_{\mu\mu'\nu\nu'}(\omega) = \nonumber \\
= \sum\limits_{n>0} \left(\begin{array}{c} {\cal X}^{n}_{\mu\mu'} \\ {\cal Y}^{n}_{\mu\mu'} \end{array}\right)
\frac{1}{\omega - \omega_n + i\delta}\left(\begin{array}{cc} {\cal X}^{n\ast}_{\nu\nu'} & {\cal Y}^{n\ast}_{\nu\nu'} \end{array}\right)\nonumber \\
- \sum\limits_{n>0} \left(\begin{array}{c} {\cal Y}^{n\ast}_{\mu\mu'} \\ {\cal X}^{n\ast}_{\mu\mu'} \end{array}\right)
\frac{1}{\omega + \omega_n - i\delta}\left(\begin{array}{cc} {\cal Y}^{n}_{\nu\nu'} & {\cal X}^{n}_{\nu\nu'} \end{array}\right),\nonumber \\
\label{Romega}
\eea
with the transition density components:
\be
{\cal X}^{n}_{\mu\mu'} = \langle 0|\alpha_{\mu'}\alpha_{\mu}|n\rangle \ \ \ \ \ \ \ \ \ \ {\cal Y}^{n}_{\mu\mu'} = \langle 0|\alpha^{\dagger}_{\mu}\alpha^{\dagger}_{\mu'}|n\rangle .
\label{XY}
\ee
From Eq. (\ref{Romega}), one deduces the time-dependent form of the superfluid response function as
\bea
{\hat{\cal R}}_{\mu\mu'\nu\nu'} (t-t') = \nonumber
\\
= -i\langle T\left(\begin{array}{cc}
(\alpha_{\mu'}\alpha_{\mu})(t)(\alpha^{\dagger}_{\nu}\alpha^{\dagger}_{\nu'})(t') & 
(\alpha_{\mu'}\alpha_{\mu})(t)(\alpha_{\nu'}\alpha_{\nu})(t') \\
(\alpha^{\dagger}_{\mu}\alpha^{\dagger}_{\mu'})(t)(\alpha^{\dagger}_{\nu}\alpha^{\dagger}_{\nu'})(t') &
(\alpha^{\dagger}_{\mu}\alpha^{\dagger}_{\mu'})(t)(\alpha_{\nu'}\alpha_{\nu})(t')
\end{array}\right)\rangle 
\nonumber\\
= -i
\langle T\left(\begin{array}{cc}
A_{\mu\mu'}(t)A^{\dagger}_{\nu\nu'}(t')  &
A_{\mu\mu'}(t)A_{\nu\nu'}(t') \\
A^{\dagger}_{\mu\mu'}(t)A^{\dagger}_{\nu\nu'}(t') &
A^{\dagger}_{\mu\mu'}(t)A_{\nu\nu'}(t')
\end{array}\right)\rangle ,\nonumber\\
\label{RtA}
\eea
implying the time-dependent operator products in the Heisenberg picture:
\bea
A_{\mu\mu'}(t) = (\alpha_{\mu'}\alpha_{\mu})(t) = e^{iHt}\alpha_{\mu'}\alpha_{\mu}e^{-iHt} \nonumber \\ 
A^{\dagger}_{\nu\nu'}(t) = (\alpha^{\dagger}_{\nu}\alpha^{\dagger}_{\nu'})(t) = 
e^{iHt}\alpha^{\dagger}_{\nu}\alpha^{\dagger}_{\nu'}e^{-iHt}.
\eea 
A Fourier transformation of Eq. (\ref{RtA}) should yield Eq. (\ref{Romega}) and vice versa to verify this. 
Here we also note that the response function is an internal characteristic of the many-body system disentangled from the strength function by Eq. (\ref{Polar}).

The EOM for the superfluid response (\ref{RtA}) is generated similarly to the EOMs for single-fermion propagators. 
Eq. (\ref{RtA}) is differentiated sequentially with respect to $t$ and $t'$ and, taking the Fourier image, one obtains
\bea
{\hat{\cal R}}_{\mu\mu'\nu\nu'}(\omega) = {\hat{\cal R}}^0_{\mu\mu'\nu\nu'}(\omega) \nonumber\\
+ \frac{1}{4}\sum\limits_{\gamma\gamma'\delta\delta'}{\hat{\cal R}}^0_{\mu\mu'\gamma\gamma'}(\omega){\hat{\cal K}}_{\gamma\gamma'\delta\delta'}(\omega){\hat{\cal R}}_{\delta\delta'\nu\nu'}(\omega),
\label{BSDE}
\eea
which is the Bethe-Salpeter equation but simplified to the Dyson form since the two-point CF is considered from the beginning.
We will refer to it as Bethe-Salpeter-Dyson equation (BSDE).
Eq. (\ref{BSDE}) has, however, in addition, the $2\times 2$ matrix structure in the quasiparticle basis. The free (uncorrelated) response is defined as
\be
{\hat{\cal R}}^0_{\mu\mu'\nu\nu'}(\omega) = \left[\omega\right. - \left.{\hat\sigma}_3E_{\mu\mu'}\right]^{-1}{\hat{\cal N}}_{\mu\mu'\nu\nu'},
\label{R0}
\ee
with
\be
E_{\mu\mu'} = E_{\mu} + E_{\mu'}, \ \ \ \ \ \ \ \ \ \ \ \ \ \ \ {\hat\sigma}_3 = \left(\begin{array}{cc} 1& 0 \\ 0 & -1 \end{array}\right),
\ee 
and the norm matrix ${\hat{\cal N}}_{\mu\mu'\nu\nu'}$, which is specified below.
The static and dynamical parts of the interaction kernel ${\hat{\cal K}}(\omega) = {\hat{\cal K}}^0 + {\hat{\cal K}}^r(\omega)$ read:
\bea
{\hat{\cal K}}^0_{\gamma\gamma'\delta\delta'} = \frac{1}{4}\sum\limits_{\eta\eta'\rho\rho'}{\hat{\cal N}}^{-1}_{\gamma\gamma'\eta\eta'}
{\hat{\cal T}}^{0}_{\eta\eta'\rho\rho'} 
{\hat{\cal N}}^{-1}_{\rho\rho'\delta\delta'} \nonumber\\
{\hat{\cal K}}^r_{\gamma\gamma'\delta\delta'}(\omega) = \frac{1}{4}\sum\limits_{\eta\eta'\rho\rho'}\left[{\hat{\cal N}}^{-1}_{\gamma\gamma'\eta\eta'}
{\hat{\cal T}}^{r}_{\eta\eta'\rho\rho'}(\omega) 
{\hat{\cal N}}^{-1}_{\rho\rho'\delta\delta'}\right]^{irr},\nonumber\\
\label{Kernel}
\eea
where the static and time-dependent ${\hat{\cal T}}$-matrices in the quasiparticle space in the most general form are:
\be
{\hat{\cal T}}^{0}_{\mu\mu'\nu\nu'} = -\langle\left(\begin{array}{cc}\left[[V,A_{\mu\mu'}],A^{\dagger}_{\nu\nu'}\right]  & \Bigl[[V,A_{\mu\mu'}],A_{\nu\nu'}\Bigr] 
\\
\\
\left[[V,A^{\dagger}_{\mu\mu'}],A^{\dagger}_{\nu\nu'}\right]  &
\left[[V,A^{\dagger}_{\mu\mu'}],A_{\nu\nu'}\right]
\end{array}\right)\rangle
\label{T02}
\ee
\bea
{\hat{\cal T}}^{r}_{\mu\mu'\nu\nu'}(t-t') = i\times\ \ \ \ \ \ \ \ \ \ \ \ \ \ \ \ \ \ \ \ \ \ \ \ \ \ \ \ \ \ \ \ \ \ \ \ \ \ \ \ \ \ \ \ \ \
\nonumber\\ \times
\langle T\left(\begin{array}{cc}[V,A_{\mu\mu'}](t)[V,A^{\dagger}_{\nu\nu'}](t')  & [V,A_{\mu\mu'}](t)[V,A_{\nu\nu'}](t')
\\
\left[V,A^{\dagger}_{\mu\mu'}\right](t)[V,A^{\dagger}_{\nu\nu'}](t')  &
[V,A^{\dagger}_{\mu\mu'}](t)[V,A_{\nu\nu'}](t')
\end{array}\right)\rangle . \nonumber\\
\label{Tr2}
\eea
The norm matrix ${\hat{\cal N}}_{\mu\mu'\nu\nu'}$ becomes, accordingly:
\be
{\hat{\cal N}}_{\mu\mu'\nu\nu'} = \langle\left(\begin{array}{cc}[A_{\mu\mu'},A^{\dagger}_{\nu\nu'}]  & 0
\\
0  &
[A^{\dagger}_{\mu\mu'},A_{\nu\nu'}]
\end{array}\right)\rangle ,
\label{norm}
\ee
while its inverse is given by
\bea
\frac{1}{2}\sum_{\delta\delta'}{\hat{\cal N}}^{-1}_{\mu\mu'\delta\delta'}{\hat{\cal N}}_{\delta\delta'\nu\nu'} = \delta_{\mu\mu'\nu\nu'} =
 \delta_{\mu\nu}\delta_{\mu'\nu'} - \delta_{\mu\nu'}\delta_{\mu'\nu}.\nonumber\\
\eea

The superfluid response theory is thus comprised of the BSDE (\ref{BSDE}) with the uncorrelated quasiparticle propagator (\ref{R0}) and interaction kernel (\ref{Kernel}) in their most general forms. At this point, the theory is 
still exact, but too general. To proceed further, the evaluation of the commutators figuring in Eqs. (\ref{T02}, \ref{Tr2}, \ref{Tr2}) is required. 
The exact forms of the $ph$ and $pp$ static kernels separately were presented and discussed, e.g., in Refs. 
\cite{SchuckTohyama2016a,LitvinovaSchuck2019,Schuck2021}. Both contain the pure contribution of the bare fermionic interaction and its contractions with the correlated parts of the two-body fermionic densities. It is further pointed out that the latter densities are static limits of the two-fermion propagators, so that the correlated self-consistent static kernel contains feedback from the dynamical kernel. The superfluid analogs of these terms were not yet presented in the literature, and I will discuss them elsewhere.
The ab initio superfluid static kernel is the unification of the $ph$ and $pp$ static kernels which form the main diagonal of the block matrix (\ref{T02}). Dropping the correlated terms is technically confining by the HFB approximation. In this case, the superfluid static kernel simplifies to the kernel of QRPA. Its matrix elements are determined by the following expectation values \cite{RingSchuck1980}: 
\bea
\langle\text{HFB}|\left[[V,A_{\mu\mu'}],A^{\dagger}_{\nu\nu'}\right]|\text{HFB}\rangle 
&=& -H^{22}_{\mu\mu'\nu\nu'},\nonumber
\\
\langle\text{HFB}|\left[[V,A_{\mu\mu'}],A_{\nu\nu'}\right]|\text{HFB}\rangle &=& 4!H^{40}_{\mu\mu'\nu\nu'},\nonumber
\\
{\hat{\cal N}}_{\mu\mu'\nu\nu'} &=& {\hat{\sigma}}_3\delta_{\mu\mu'\nu\nu'},
\nonumber\\
\label{ABN}
\eea
and the remaining matrix elements of Eq. (\ref{T02}) can be determined via Hermitian conjugation. 
\begin{figure*}
\begin{center}
\includegraphics[scale=0.70]{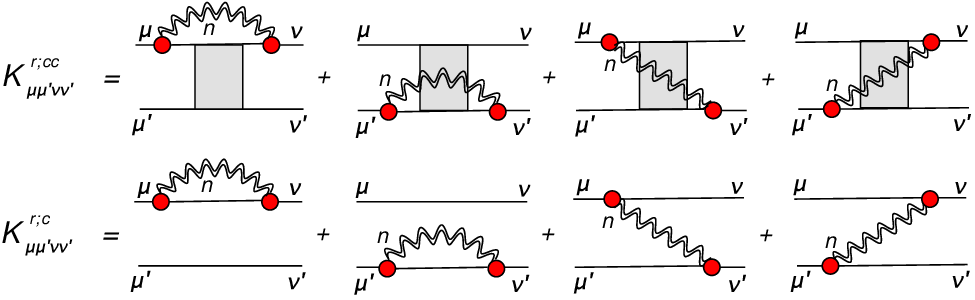}
\end{center}
\caption{The superfluid dynamical two-fermion kernels in the qPVC approximation. The complete multicomponent two-quasiparticle correlation function (\ref{RtA}) is denoted by the shaded rectangular block. Top: the kernel of Eq. (\ref{Kr11cc}) with two two-quasiparticle correlation functions; bottom: the kernel of Eq. (\ref{Kr11c}) with one two-quasiparticle correlation function, or leading-order qPVC.}
\label{Kr_qvc}%
\end{figure*}

The dynamical kernel of Eq. (\ref{Tr2}) can be evaluated similarly and, in general, is quite complicated. The dominant contribution to the excited states comes from the diagonal components related as follows:
\be
{\cal K}^{r[22]}_{\mu\mu'\nu\nu'}(\tau) = {\cal K}^{r[11]}_{\nu\nu'\mu\mu'}(-\tau),
\ee
so that it is sufficient to obtain explicitly only ${\cal K}^{r[11]}$. The non-diagonal terms are associated with complex ground state correlations, playing a relatively minor role in the gross features of the giant resonances. However, they may become important in quantifying fine spectral details, especially at low energies below the particle emission threshold, where the density of states is low.

Here I focus on ${\cal K}^{r[11]}$ as an example of how to proceed that can be applied to all components of the dynamical kernel.
Its $T$-matrix reducible precursor
\be
{\cal T}^{r[11]}_{\mu\mu'\nu\nu'}(t-t') = i\langle T[V,A_{\mu\mu'}](t)[V,A^{\dagger}_{\nu\nu'}](t') 
\rangle
\label{Tr11}
\ee
is a time-ordered product of eight quasiparticle creation and destruction operators, four at time $t$ and four at time $t'$. This constitutes a fully correlated two-times four-quasiparticle propagator contracted with two matrix elements of $\bar v$; see Ref.  \cite{Litvinova2022} for details. 
Through this CF in the dynamical kernel, the two-fermion response ${\hat{\cal R}}$ couples to EOMs for growing-rank CFs, similarly to the single-quasiparticle propagator, and again, various approximations may be generated by a cluster decomposition analogous to the one of Eq. (\ref{CD}):
\bea
G^{(4q)irr} \sim G^{(q)}G^{(q)}G^{(q)}G^{(q)} + G^{(q)}G^{(q)}G^{(2q)} \nonumber \\
+ G^{(2q)}G^{(2q)} + G^{(q)}G^{(3q)} + \sigma^{(4q)},
\label{CD2}
\eea
where the symbol $G^{(nq)}$ stands for any type of correlation functions with $n$ quasiparticle operators. 
The qPVC approaches are associated with retaining the terms with at least one $G^{(2q)}$ which, by contraction with the pairs of interaction matrix elements, can be exactly mapped to the qPVC amplitudes as in Eqs. (\ref{mappingph}, \ref{mappingpp}, \ref{Gamma11_qp}, \ref{Gamma02_qp}).

Here, I narrow the discussion to the qPVC factorizations, which keep all the possible $G^{(2q)}$ contributions. As in the case of the single-fermion self-energy, the terms with $G^{(q)}G^{(q)}$ are partly absorbed in the $G^{(2q)}$ CFs, while their remaining contributions should be small in the leading-order qPVC at strong coupling. They can be evaluated afterward if better accuracy is sought.
The leading contribution to ${\cal K}^{r[11]}$ takes the form \cite{Litvinova2022}:
\bea
{\cal K}^{r[11]cc}_{\mu\mu'\nu\nu'}(\omega) &=& 
\sum\limits_{\gamma\delta nm}\Bigl[\frac{\Gamma^{(11)n}_{\mu\gamma}{\cal X}^{m}_{\mu'\gamma}{\cal X}^{m\ast}_{\nu'\delta}\Gamma^{(11)n\ast}_{\nu\delta}}{\omega - \omega_{nm} + i\delta} \nonumber\\ &-& 
\frac{\Gamma^{(11)n\ast}_{\gamma\mu}{\cal Y}^{m\ast}_{\mu'\gamma}{\cal Y}^{m}_{\nu'\delta}\Gamma^{(11)n}_{\delta\nu}}{\omega + \omega_{nm} - i\delta}\Bigr]
- \cal{AS},
\label{Kr11cc}
\eea
where $\omega_{nm} = \omega_{n} + \omega_{m} $ and $\cal{AS}$ stands for antisymmetrizations, and the upper index "cc" indicates the approximation retaining CFs up to $G^{(2q)}G^{(2q)}$. The top line of Fig. \ref{Kr_qvc} gives the diagrammatic interpretation of Eq. (\ref{Kr11cc}).
The two-quasiparticle CFs figuring in the dynamical kernel (\ref{Kr11cc}) in the form of $\{{\cal X}, {\cal Y}\}$ components and frequencies $\omega_{nm}$ are formally exact and, thus, are not associated with any particular approximations, such as perturbative expansions or partial resummations. A remarkable consequence of this observation is that the cluster approaches retaining $G^{(2q)}$ CFs can include arbitrarily complex 2n-quasiparticle configurations non-perturbatively, which can be generated, e.g., iteratively in a self-consistent cycle. On the other hand, various approximations can certainly be applied to calculations of these CFs and associated phonon characteristics.

As in the non-superfluid case, only one of the two two-fermion CFs can form a phonon, because only two matrix elements of the NN interaction are present in Eq. (\ref{Tr2}). Because of that, only one CF contracted with these matrix elements can be mapped to the qPVC. 
The correlations in the other two-fermion CF of Eq. (\ref{Kr11cc}), which is not associated with a phonon but rather forms an intermediate propagator (rectangular block in Fig. \ref{Kr_qvc}), can be relaxed, which leads to the qPVC-NFT approximation:
\bea
{\cal K}^{r[11]c}_{\mu\mu'\nu\nu'}(\omega) = 
\Bigl\{\Bigl[ \delta_{\mu'\nu'}
\sum\limits_{\gamma n}\frac{\Gamma^{(11)n}_{\mu\gamma}\Gamma^{(11)n\ast}_{\nu\gamma}}{\omega - \omega_{n} - E_{\mu'} - E_{\gamma}}  \nonumber\\ - 
\sum\limits_{n}\frac{\Gamma^{(11)n}_{\mu\nu'}\Gamma^{(11)n\ast}_{\nu\mu'}}{\omega - \omega_{n} - E_{\mu'} - E_{\nu'}}\Bigr]
- \Bigl[ \mu\leftrightarrow\mu'\Bigr]\Bigr\} - \Bigl\{ \nu\leftrightarrow\nu'\Bigr\}.\nonumber\\
\label{Kr11AAa_qPVC_0}
\eea
The index "c" marking this version of the dynamical kernel signals that only one two-quasiparticle CF is retained.
Eq. (\ref{Kr11AAa_qPVC_0}) can be further transformed to the form recognizable as a superfluid generalization of the NFT's dynamical kernel by rearranging the antisymmetrizations as follows:
\bea
{\cal K}^{r[11]c}_{\mu\mu'\nu\nu'}(\omega) =  \ \ \ \ \ \ \ \ \ \ \ \ \ \ \ \ \ \ \ \  \nonumber\\
= \Bigl[ \delta_{\mu'\nu'}
\sum\limits_{\gamma n}\frac{\Gamma^{(11)n}_{\mu\gamma}\Gamma^{(11)n\ast}_{\nu\gamma}}{\omega - \omega_{n} - E_{\mu'\gamma}}  
+\delta_{\mu\nu}
\sum\limits_{\gamma n}\frac{\Gamma^{(11)n}_{\mu'\gamma}\Gamma^{(11)n\ast}_{\nu'\gamma}}{\omega - \omega_{n} - E_{\mu\gamma}}
\nonumber\\ + 
\sum\limits_{n}\frac{\Gamma^{(11)n}_{\mu\nu}\Gamma^{(11)n\ast}_{\nu'\mu'}}{\omega - \omega_{n} - E_{\mu'\nu}}
+ \sum\limits_{n}\frac{\Gamma^{(11)n}_{\mu'\nu'}\Gamma^{(11)n\ast}_{\nu\mu}}{\omega - \omega_{n} - E_{\mu\nu'}}\Bigr] \nonumber\\
- \Bigl[ \nu\leftrightarrow\nu'\Bigr].\ \ \ \ \ \ \ \ \ \ \ \ 
\label{Kr11c}
\eea
If the pairing correlations are described by the BCS approximation, the dynamical kernels (\ref{Kr11cc}, \ref{Kr11c}) have the same form, but the quasiparticle energies $E_{\mu}$ and qPVC vertex functions $\Gamma^{(ij)n}_{\mu\nu}$ are replaced by their BCS analogs, where the $U,V$ matrices are diagonal  \cite{Zelevinsky2017}.
The resulting kernel (\ref{Kr11c}) is analogous to the resonant superfluid dynamical kernels of the NFT \cite{Niu2016} and quasiparticle time blocking approximation \cite{Tselyaev2007,LitvinovaTselyaev2007} where the qPVC was introduced phenomenologically based on the assumption of the existence of the phonon modes and their coupling to quasiparticles within the second-order perturbation theory. As mentioned above,  Eqs. (\ref{Kr11cc}, \ref{Kr11c}) correspond to the leading resonant approximations to the dynamical kernel, where a number of complex ground state correlations were neglected.  

\subsection{qPVC approaches for nuclear response in a relativistic self-consistent framework}

The self-consistent implementations of the nuclear response theory beyond QRPA employed DFT-based effective in-medium NN interactions and predominantly explored the resonant qPVC dynamical kernel ${\cal K}^{r[11]c}$.  The Skyrme interactions of zero-range were used, e.g., in Refs. \cite{Tselyaev2018,Lyutorovich2018,Niu2016,Niu2018}, and the QRPA+qPVC based on the relativistic meson-exchange interaction has been available since Ref. \cite{LitvinovaRingTselyaev2008}. The latter method was formulated analytically in terms of the relativistic time blocking approximation (RQTBA) starting from the Bethe-Salpeter equation for the four-time two-quasiparticle propagator with the phenomenologically inputed qPVC, following its original non-relativistic version \cite{Tselyaev1989,Tselyaev2007,LitvinovaTselyaev2007}. The resonant superfluid dynamical kernels of Refs. \cite{Tselyaev2007,LitvinovaTselyaev2007,LitvinovaRingTselyaev2008} are consistent with the kernel of Eq. (\ref{Kr11c}) in the BCS limit. However, we note that the time blocking and the necessity of the phenomenological qPVC can be ruled out in the ab initio theory discussed here. RQTBA, representing the two-fermion sector of the relativistic nuclear field theory (RNFT), demonstrated good performance for the neutral \cite{LitvinovaLoensLangankeEtAl2009,LitvinovaRingTselyaev2010,EndresLitvinovaSavranEtAl2010,MassarczykSchwengnerDoenauEtAl2012, LanzaVitturiLitvinovaEtAl2014,PoltoratskaFearickKrumbholzEtAl2014,NegiWiedekingLanzaEtAl2016,EgorovaLitvinova2016,Carter2022,Litvinova2023,Markova2024,Markova2025} and charge-exchange \cite{RobinLitvinova2016,Scott2017,RobinLitvinova2018,Robin2019} excitations describing a large variety of nuclear phenomena.

In RQTBA, the inclusion of the qPVC dynamical kernel already in the leading order with respect to the qPVC vertex (\ref{Kr11c}) showed significant improvements as compared to QRPA. The most notable effect is seen on the widths of the giant resonances, first of all, the primarily studied ones with zero and one units of the angular momentum transfer: monopole, dipole, Gamow-Teller, and spin-dipole resonances. At the same time, the low-energy parts of the respective spectra are populated by crowds of fragments of the "primordial" QRPA soft modes and partly the giant resonances, also improving the agreement with data. The associated characteristics, such as beta decay rates, nuclear compressibility, and other nuclear structure properties, are improved accordingly.  Remarkably, RNFT enabled such a description in a single framework across the nuclear chart with only input from the covariant DFT \cite{Ring1996,VretenarAfanasjevLalazissisEtAl2005,Lalazissis1997,NL3star} and essentially the same parameter set.
The self-consistency put stringent constraints on the calculation schemes via (i) approximating the static kernels of the one-fermion and two-fermion EOMs by the first and second variational derivatives of the energy density functional, respectively, (ii) obtaining the phonon characteristics with the same static kernels, and (iii) the subtraction of the static limit of the dynamical kernel \cite{Tselyaev2013} to eliminate the double counting of qPVC contained implicitly in the effective interaction.    
The complete self-consistency, the covariance of RNFT, and its rooting in particle physics provide an optimal balance of fundamentality, accuracy, and feasibility. The theory appears as predictive, transferrable across the energy scales, and systematically improvable due to the variability of the dynamical kernel and its organization in accordance with the qPVC power counting. The latter two features were enabled after the completion of the ab initio EOM qPVC framework \cite{LitvinovaSchuck2019,LitvinovaSchuck2020,Litvinova2021a,Litvinova2022} outlined in these lectures. Further effort allowed a formulation of RNFT for finite temperatures \cite{LitvinovaWibowo2018,LitvinovaWibowo2019,LitvinovaRobinWibowo2020,Litvinova2021b} extending its applications to astrophysically relevant phenomena.

\section{Nuclear response in the RNFT framework}
\label{calculations}
\subsection{Electromagnetic response}

The electromagnetic response is the most studied type of nuclear response as it can be induced by the most accessible experimental probes with photons \cite{Ishkhanov2021,GR2001,SavranAumannZilges2013}. In theory, excitation operators are assigned to those probes according to their transferred angular momentum $L$ and parity $\pi$ \cite{GR2001}.
The electric operators are associated with natural parity when $\pi = (-1)^L$: 
\bea
F_{00} &=& e\sum\limits_{i = 1}^Zr_i^2,\nonumber \\
F_{1M} &=& \frac{eN}{A}\sum\limits_{i=1}^Z r_iY_{1M}({\hat{\bf r}}_i) - \frac{eZ}{A}\sum\limits_{i=1}^N r_iY_{1M}({\hat{\bf r}}_i),
\nonumber\\
F_{LM} &=& e\sum\limits_{i = 1}^Zr_i^{\ L}Y_{LM}({\hat{\bf r}}_i),\qquad L \geq 2.
\label{FEL}
\eea
In Eqs. (\ref{FEL}) $e$ is the positron charge, $Y_{LM}({\hat{\bf r}})$ are the spherical harmonics, and $N$ and $Z$ are the neutron and proton numbers in the nucleus under study, respectively. The dipole operator $F_{1M}$ is corrected for the center of mass motion by introducing the ``kinematic'' charges so that it also engages the neutron subsystem. The electric excitation operators of other multipolarities couple to proton subsystems only. This operator type is classified as the isovector one with $\Delta T = 1$, i.e., change of the isospin by one unit.

Complementary probes, e.g., hadronic ones, are related to isoscalar operators. The latter contain summations over all the nucleons and do not cause isospin transfer. By their nature, isoscalar operators couple to protons and neutrons in the same way, and no electric charge is involved, i.e., one sets $e =1$ in Eqs. (\ref{FEL}) for $L \neq 1$. 
The isoscalar dipole operator is defined as an overtone of the dipole operator, or next to leading order multipole expansion: 
\be
F^{(IS)}_{1M} = \sum\limits_{i=1}^A (r^3_i - \eta r_i)Y_{1M}({\hat{\bf r}}_i),  
\label{opISE1}
\ee 
which is reflected in the cubic radial dependence. The reason is that the linear radial dependence of the isoscalar dipole operator only induces a translation of the nucleus as a whole and does not produce internal excitations.
The second term in Eq. (\ref{opISE1}), $\eta = 5\langle r^2\rangle/3$, is introduced to eliminate the translational mode (spurious) admixture to the excitation spectrum 
\cite{GR2001,Garg2018}. Isoscalar operators are assigned $\Delta T = 0$ with no change of isospin.

The operators of the unnatural parity $\pi = (-1)^{L+1}$ are associated with magnetic multipole transitions with a more complex structure involving spin transfer. 
Magnetic transitions are not as collective as electric ones \cite{Tselyaev2020}. The underlying reason for reduced collectivity is that the leading spin-transfer part of the NN interaction is mediated by single pion and rho-meson exchange, whose vertices contain smallness of $1/M_N$, where $M_N$ is the nucleon mass, compared to the non-spin-flip meson vertices \cite{Boyussy1987}.
I will focus on the electric dipole transitions in this subsection, leaving magnetic excitations to a separate discussion. Some studies of magnetic resonances can be found in Refs. \cite{Tselyaev2020,KamerdzhievSpethTertychny2004,Oishi2019}.

The schematic model by Brown and Bolsterli \cite{BrownBolsterli1959,RingSchuck1980} implying the equidistant particle-hole states and separability of the multipole-multipole interaction reveals generic features of the excitation spectra of strongly correlated systems on the RPA level. 
The spectrum starts and ends with two highly collective states, the low-frequency and the high-frequency ones, respectively. Their collectivity is stipulated by coherent contributions from the uncorrelated $ph$-excitations with respect to the Fermi energy, enforced by the interparticle interaction. The other $ph$ states lie between the two collective solutions and are mostly non-collective. In more realistic RPA calculations, the resulting spectrum is sensitive to the residual interaction and to the model space size accommodated by the numerical implementation. The general composition of the spectrum inherits the gross structure of the Brown-Bolsterli model, but the two major collective solutions undergo fragmentation associated with the Landau damping in the language of the Fermi liquid theory. In practice, the low-energy solutions (soft modes) in medium-heavy nuclei are less collective than the high-energy ones (giant resonances) in the $L = 0$ and $L =1$ channels but show enhanced collectivity in the quadrupole channel, especially in axially deformed systems.
The giant resonances are classified as collective oscillations involving practically all the nucleons. 
In approximations beyond RPA taking into account the dynamical kernel in any of the forms discussed above, the presence of poles generates further fragmentation of the $ph$ states, and this effect is attributed to their coupling to more complex configurations. The degree of fragmentation beyond the Landau damping and fine details of the obtained spectra are sensitive to the approximation employed for $K^{(r)}$. 

\begin{figure*}[t]
\begin{center}
\includegraphics[width=0.75\textwidth]{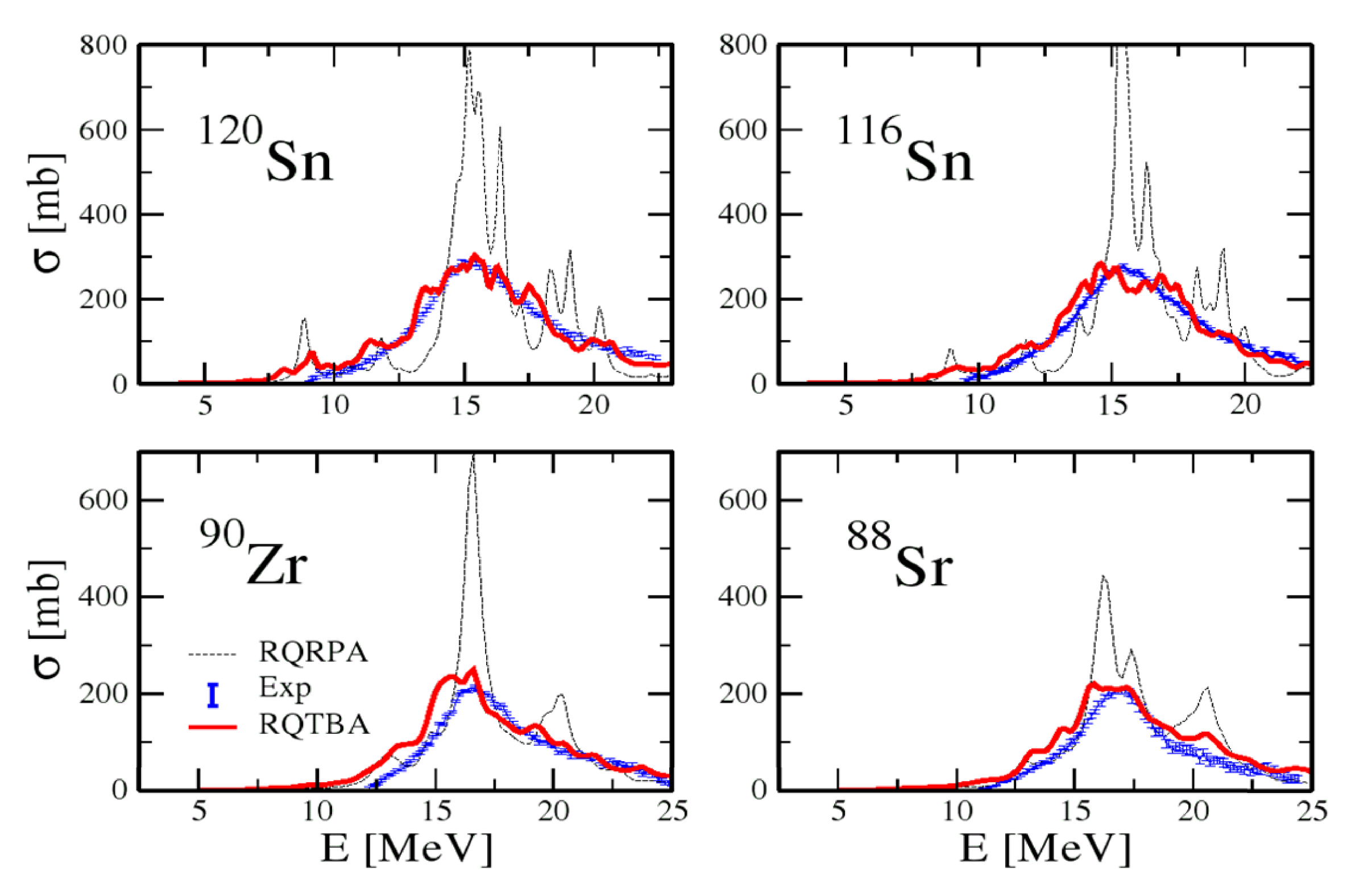}
\end{center}
\vspace{-0.3cm}
\caption{Total dipole photoabsorption cross section in stable medium-mass nuclei. The figure is adapted from Refs. \cite{50BCS,Meng2016}.}
\label{gdr}%
\end{figure*}
An illustration of the most studied dipole spectra in selected medium-mass spherical nuclei, characterized by a broad and relatively smooth peak, known as giant dipole resonance (GDR) \cite{Ishkhanov2021}, is displayed in
Figure~\ref{gdr}. The measurements are typically characterized by the total dipole photoabsorption cross section related to the strength function as follows:
\be
\sigma_{E1}(E)={\frac{{16\pi^{3}e^{2}}}{{9\hbar c}}}E~S_{E1}(E).
\label{cs}
\ee
In the examples of Figure~\ref{gdr}, 
relativistic QRPA (RQRPA)\cite{PaarRingNiksicEtAl2003} (black dashed curves) and RQTBA \cite{LitvinovaRingTselyaev2008} (red solid curves) are compared with the evaluated data (blue error bars) from the EXFOR library~\cite{nndc}. The NFT form (\ref{Kr11c}) of the qPVC dynamical kernel was used in RQTBA, i.e., the leading-order qPVC contribution was included. Finding the static kernel of the response EOM, sometimes called the bosonic mean field, from the bare NN interaction is a highly non-trivial task requiring an iterative procedure for the feedback from the dynamical kernel and removal of the confining potential artifacts. While a full ab initio calculation scheme of this kind remains a task for the future, using an effective interaction in both kernel components appears as an optimal variant.
In RQTBA, we used the effective meson-exchange interaction adjusted to nuclear masses and radii in the framework of the covariant DFT \cite{VretenarAfanasjevLalazissisEtAl2005, Meng2006, Meng2016} with the NL3 parametrization \cite{Lalazissis1997}. In a self-consistent calculation scheme with such interactions, the RQRPA generally produces the GDR, which is mostly concentrated in a narrow energy region. Its centroid and the total strength are reproduced fairly reasonably as well as the energy-weighted sum rule (EWSR), also known as the Thomas-Reiche-Kuhn (TRK) sum rule:
\be
S_{E1} = \sum\limits_{\nu} E_{\nu}B_{\nu} = \frac{9\hbar^2e^2}{8m_p}\frac{NZ}{A}.
\label{sr}
\ee
The right hand side of Eq.~(\ref{sr}) is evaluated via a double commutator of the dipole excitation operator and the nuclear Hamiltonian approximated by the non-velocity-dependent interaction \cite{RingSchuck1980}. The relation~(\ref{sr}) is, therefore, referred to as the "model-independent" sum rule. However, realistic energy density functionals (EDFs), such as the Skyrme, Gogny, and relativistic ones are based on effective interactions, which contain some dependencies on the nucleonic momenta. Therefore, in practice, the integrated theoretical spectra yield $10$--$20\%$ or even larger enhancement of the TRK EWSR in the (Q)RPA calculations \cite{Trippa2008}. This is in agreement with data spaning wide energy intervals \cite{nndc}. 

The singular dynamical kernels compatible with locality and unitarity
conserve the EWSR and some non-energy-weighted sum rules \cite{Tselyaev2007}, which serves as a helpful test for numerical implementations of approaches beyond QRPA. In particular, the resonant qPVC kernels of the NFT form have this property. 
Implementations using effective interactions to approximate the entire static kernel need an additional correction if the effective interaction is adjusted on the mean-field level as it is done in the DFT. In this case, it is implicitly assumed that the mean field stands for the entire interaction kernel so that the dynamical part is included in it in the static limit. Therefore, when the dynamical kernel is included explicitly in such calculation schemes, its static limit is subtracted as it was formulated in Ref. \cite{Tselyaev2013}. This allows one to eliminate the double counting of the qPVC effects from the effective interactions of EDFs. In practice, the subtraction is a simple procedure of the replacement
\bea
{\tilde{\cal K}}^{0} + {\tilde{\cal K}}^{r}(\omega) \to {\tilde {\cal K}}^{0}  + \delta {\tilde {\cal K}}^{r}(\omega) \nonumber \\
= {\tilde{\cal  K}^{0}}  + {\tilde {\cal K}}^{r}(\omega) - {\tilde {\cal K}}^{r}(0), 
\label{subtraction}
\eea
that is, the dynamical kernel in the static approximation $\omega = 0$ is subtracted from the dynamical kernel itself. Thus, the energy-independent combination ${\tilde {\cal K}}^{0} - {\tilde{\cal K}}^{r}(0)$ plays the role of effective interaction freed from the long-range effects of ${\tilde{\cal K}}^{r}(\omega)$, i.e., corresponds to the "refined" static kernel. In Eq.~(\ref{subtraction}), the `$\ \widetilde{ }$\ ' sign denotes the kernels, where the effective interaction is used.  The subtraction (\ref{subtraction}) 
causes some violation of the EWSR by modifying the constant part of the interaction kernel. As a result, the GDR's centroid is pushed upward as compared to qPVC calculation without subtraction, so that the final position of the major peak comes close to its (Q)RPA placement. The qPVC dynamical kernel alone (without the subtraction) shifts the GDR's main peak to lower energies. This can be a desirable feature for the ab initio implementations as it was found in the calculations within the second RPA discussed in Ref.~\cite{PapakonstantinouRoth2009}. 

Another important observation in Figure~\ref{gdr} is a notable fragmentation leading to the broadening of the GDR due to the qPVC included within RQTBA. When a large number of the phonon modes is included in the dynamical kernel, the cross section takes nearly a Lorentzian shape while the irregularities of the RQRPA spectrum are suppressed. Here I stress that small values of the smearing parameter (the imaginary part of the energy variable) were used in these calculations: $\Delta = 200$~keV for the Sn isotopes and $\Delta = 400$~keV for Sr and Zr in both the RQRPA and RQTBA cases. Typically, the choice of these parameters is stipulated by matching the experimental energy resolution and estimated continuum (particle escape) width, if the continuum is not included explicitly. 
The latter contributes to the width of the high-frequency resonances at the energies above the particle emission threshold, whose typical value is $\sim 7$--$10$~MeV in stable medium-mass and heavy nuclei.  The loosely bound nuclei with a large excess of one type of nucleon have lower separation energies and may be sensitive to the coupling to the continuum states. This can be taken into account in the (Q)RPA and beyond-(Q)RPA calculations as proposed in Ref.~\cite{ShlomoBertsch1975} for RPA,  Ref. \cite{Tselyaev2016,Kamerdzhiev1998,HaginoSagawa2001,Matsuo2002,KhanSandulescuGrassoEtAl2002,Daoutidis2009} for QRPA, and Ref. \cite{LitvinovaTselyaev2007} for QRPA+qPVC by computing the (Q)RPA propagator and its EOM in the coordinate-space representation. The qPVC kernel can be transformed to the coordinate space via the single-particle wave functions as in Ref.~\cite{LitvinovaTselyaev2007}. Ref.~\cite{Tselyaev2016} proposed a modification of this method implemented in the discrete basis of the single-particle states with the box boundary conditions. The core of these methods is constructing the mean-field propagator from the regular and irregular single-particle wave functions with the Coulomb or free asymptotics.  

The continuum width is quantitatively unimportant for highly excited states of well-bound medium-mass and heavy nuclei: a typical continuum width is 
$\sim100$~keV for a single peak in the spectrum above the particle threshold (to be compared to the large spreading width of the GDR, which amounts to a few MeV). However, the role of continuum increases dramatically in light and loosely bound nuclei, see some examples in Ref.~\cite{Tselyaev2016}. At the energies above the single-nucleon emission threshold, other emission channels open, enabling another single-nucleon or multinucleon evaporation within the GDR's energy range. The inclusion of a multiparticle continuum in approaches beyond QRPA remains a desired feature of the giant resonance theory.
%
\begin{figure*}[t]
\begin{center}
\includegraphics[width=0.7\textwidth]{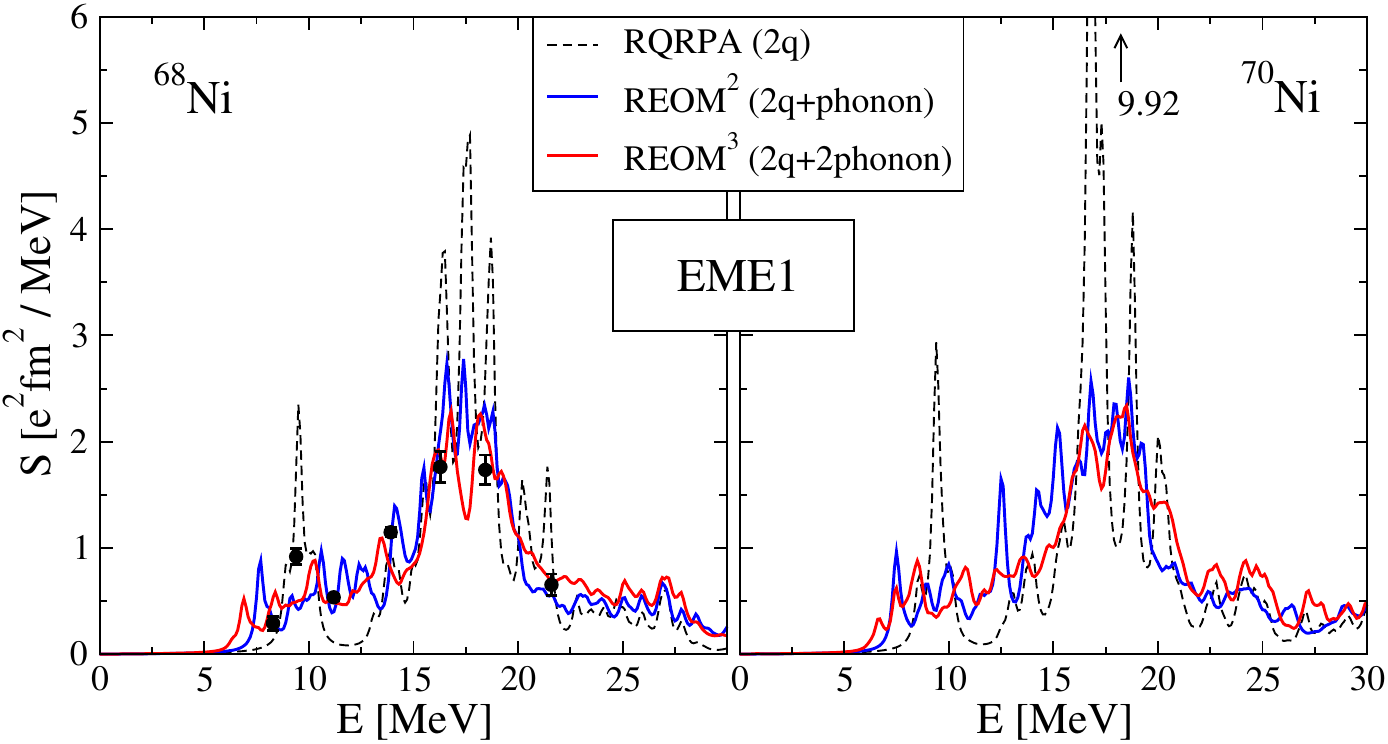}
\end{center}
\caption{Giant dipole resonance in $^{68,70}$Ni calculated within the RQRPA and REOM$^{(2,3)}$ approaches \cite{LitvinovaSchuck2019}, in comparison with the experimental data \cite{Rossi2013}. The figure is adapted from Ref. \cite{Litvinova2023a}.}
\label{68-70-ni_gdr}%
\end{figure*}
%

The models with dynamical kernels confined by correlated or non-correlated $2p2h$ configurations are overall of better quality than the QRPA ones but do not reach the spectroscopic accuracy. The latter implies the accuracy of a few hundred keV for the description of excitation spectra and associated decay properties of medium-mass and heavy nuclei, which is desirable for supporting advanced experimental high-resolution measurements and nuclear structure applications. It is clear that improvements can be made in both the EDF parametrizations and many-body methods. The majority of nuclear response calculations beyond (Q)RPA include up to the (correlated) $2p2h$ configurations \cite{LitvinovaRingTselyaev2008,LitvinovaRingTselyaev2010,NiuNiuColoEtAl2015,Gambacurta2015,RobinLitvinova2016,Tselyaev2016,Robin2019}, while calculations on the $3p3h$ level gradually become available \cite{LoIudice2012,Lenske:2019ubp,Ponomarev1999,Savran2011,Litvinova2023a,LitvinovaSchuck2019,Muescher2024}. It is instructive to compare $2p2h$ and $3p3h$ calculations within the same implementation scheme. Such a comparison indicates that the inclusion of $3p3h$ configurations essentially improve the results. At the same time, their effect is smaller than the effect of the inclusion of $2p2h$ configurations beyond (Q)RPA. This suggests that higher-complexity configurations are important for achieving spectroscopic accuracy but the theory also indicates fast saturation with respect to the configuration complexity. 

Fig.~\ref{68-70-ni_gdr} illustrates calculations for the electromagnetic dipole response of $^{68,70}$Ni where $3p3h$ configurations were included in the "two quasiparticles coupled to two phonons" ($2q\otimes 2phonon$) scheme. This was achieved by implementing an iterative cycle for the dynamical qPVC kernel.
To enter the cycle, RQRPA was run for $J\leq 6$ multipoles, and the most relevant phonon modes with the largest isoscalar excited state probabilities were selected. These phonons were used for constructing the $2q\otimes phonon$ qPVC kernel, and the BSDE for the response was solved with this kernel for the $J^{\pi}$ ($J\leq 6$) channels of both parities. The obtained response functions were collected and recycled in the dynamical kernel of the BSDE for the dipole response. This approach is dubbed here as REOM$^3$ since the dynamical kernel contains up to $3p3h$ configurations organized by qPVC.
The  REOM$^3$ electromagnetic dipole strength functions (red solid curves) are plotted in Fig.~\ref{68-70-ni_gdr} in comparison with the results of RQRPA (black dot-dashed curves), REOM$^2$ (RQTBA) (blue dashed curves) and experimental data (error bars) extracted from Ref. \cite{Rossi2013}. 

The role of the $2q\otimes phonon$ configurations in the GDR's formation in calcium isotopes was studied and discussed in Ref. \cite{EgorovaLitvinova2016}. It was established that, in the RQTBA framework, these configurations generate a notable spreading due to the fragmentation of the RQRPA modes, improving the agreement to data. However, even with a nearly complete set of the  $2q\otimes phonon$ configurations, the widths of the GDR were sizeably smaller than the observed ones and the cross sections on its high-energy shoulder were underestimated.  QTBA calculations with Skyrme forces reported a similar result \cite{Tselyaev2016}.  
Thus, going beyond the leading-order qPVC in the dynamical kernels is desirable, and the $2q\otimes 2phonon$ configurations included in Ref. \cite{LitvinovaSchuck2019} in the GDR computation for calcium isotopes indicated that the increase of configuration complexity has the potential to resolve the discrepancies of RQTBA with data at least in the medium-light nuclei. Later on, REOM$^3$ has pushed the mass limits to the tin region \cite{Novak2024} and neutron-rich unstable $^{68,70}$Ni \cite{Litvinova2023a} which I invoke as an illustration here. Since the experimental data are available only for $^{68}$Ni at low energy resolution, it is difficult to conclude unambiguously whether the higher-complexity configurations improve the GDR's description here, however, this was shown to be the case for the low-energy dipole strength discussed in Ref. \cite{LitvinovaSchuck2019}. The calculations available up until now, suggest that 
the $2q\otimes 2phonon$ configurations in REOM$^3$ cause further fragmentation of the GDR and enforce the spreading of the strength to both higher and lower energies caused by the presence of the additional poles in the interaction kernel and, consequently, in the response function.

Microscopic calculations for the response of non-spherical nuclei are more challenging. The QRPA $2q$ model space in the axially deformed, i.e., two-dimensional, coordinate basis of the relevant states is much larger than that in the spherical case, which requires only one spatial coordinate, the radial distance.  The degeneracy of $j$-orbitals is lifted in axial geometry, and the total angular momentum is no longer a good quantum number.  The direct QRPA calculations for deformed nuclei, thus, require massive numerical evaluation of matrix elements of the nucleon-nucleon interaction, which is prohibitively difficult even in the DFT frameworks \cite{PenaRing2008,Peru2008,Toivanen2010}.
However, since an elegant numerical solution was proposed in Ref.~\cite{Nakatsukasa2007} in terms of the finite-amplitude method (FAM), avoiding direct computation of the interaction matrix elements, this application area began to advance quickly
 \cite{OishiKortelainenHinohara2016,Niksic2013,Kortelainen2015}. 
\begin{figure*}[ptb]
\begin{center}
\includegraphics[scale=0.55]{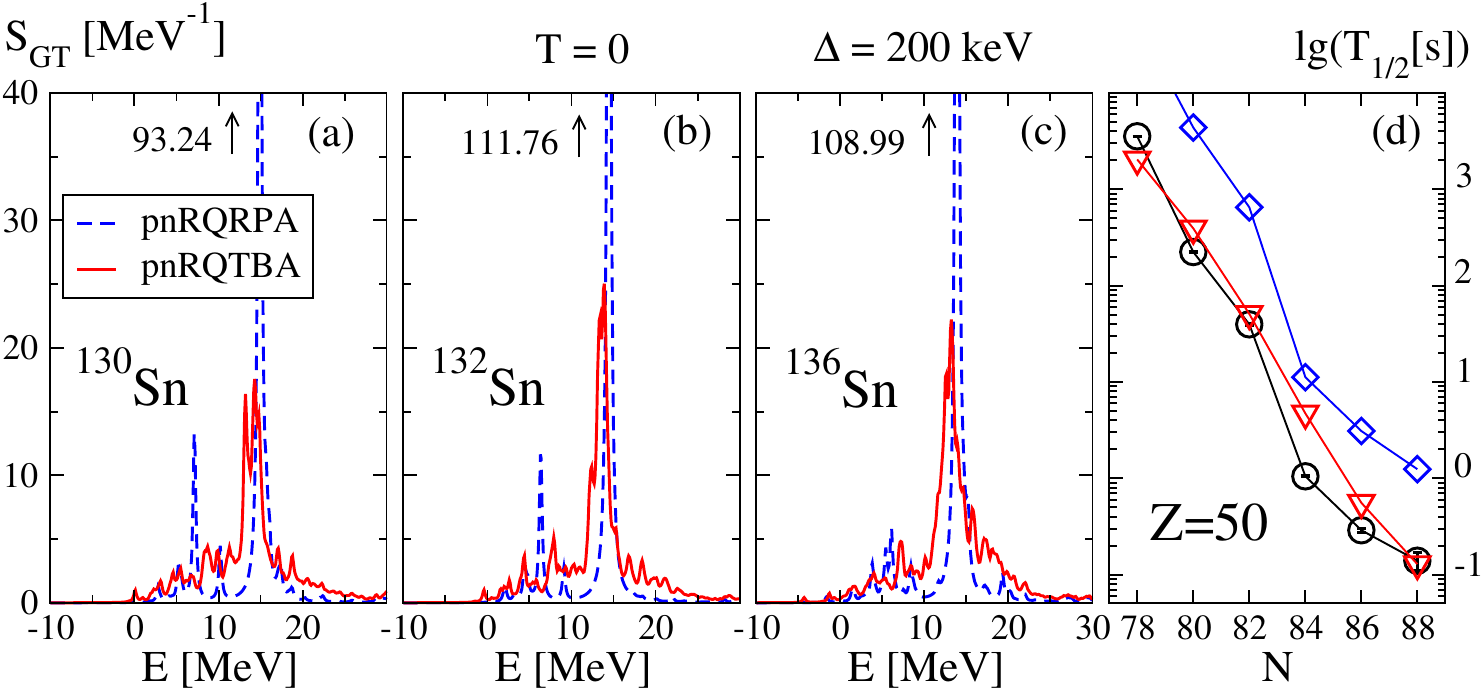}
\end{center}
\vspace{-0.2cm}
\caption{GTR in the $^{130,132,136}$Sn isotopes obtained within pnRQTBA, compared to pnRQRPA (a-c). Beta decay half-lives in the neutron-rich tin nuclei extracted from the pnRQRPA (diamonds) and pnRQTBA (triangles) GTR strength distributions, compared to experimental data (circles) \cite{nndc} (d). The figure is adapted from Ref. \cite{LitvinovaRobinWibowo2020}.}  
\label{GTR_Sn}
\end{figure*}

Formulations of the qPVC in the HFB basis \cite{Litvinova2021a,Litvinova2022} enabled a generalization of the FAM-QRPA to FAM-qPVC, with the numerical implementation in the axially deformed bases for one-fermion \cite{Zhang2022} and two-fermion \cite{Liu2024} EOMs. Such calculations showed a good performance in the description of the single-quasiparticle states and Gamow-Teller strength distributions, respectively, already in the leading-order qPVC. The description of moderately deformed nuclei can be performed in a spherical basis by generating configurations of higher complexity. Ref. \cite{Muescher2024} reported an experimental study of the dipole response of the deformed $^{64}$Ni complemented by calculations within REOM$^3$. Such calculations are possible when the low-lying phonons are accessible by spherical QRPA without encountering the instabilities associated with the deformed phase transition, which can be regulated by varying the pairing interaction strength. In this case, the self-consistently generated complex configurations capture the deformation effects dynamically instead of the static deformation imposed by the mean-field basis. 

\subsection{Spin-isospin response}

The charge-exchange resonances, which involve the conversion of a neutron to a proton and backward inside the nucleus, are associated with weak currents and decays. These excitations transfer isospin, and the most important of them are also associated with spin flip.  Thus, we move to the domain of the nuclear spin-isospin response, which consists of the transitions from some initial state (which is often the ground state) of the nucleus $(N,Z)$ to the final states in the neighboring nuclei $(N\mp1,Z\pm1)$ with the lowering $T_-$ and raising $T_+$  isospin, respectively.
Such transitions happen spontaneously in $\beta$ decays and can be induced by external perturbations, e.g., the charge-exchange nuclear reactions. Examples are $(p, n)$ and $(^3{\rm He}, t)$ reactions.
The spin-flip can occur simultaneously with the isospin transfer, so that the non-spin-flip modes ($S = 0$), and the spin-flip ones ($S = 1$) are distinguished. The most known modes are the isobaric analog state (IAS) with $S = 0$, $J^\pi = 0^{+}$, the Gamow-Teller resonance (GTR) with $S = 1$, $J^\pi = 1^{+}$, and the spin-dipole resonance (SDR) with $S = 1$, $J^\pi = 0^{-}, 1^{-}, 2^{-}$ \cite{Osterfeld1992, Ichimura2006, PaarVretenarKhanEtAl2007, Roca-Maza2018}.
The corresponding operators are assigned as follows:
\bea
F_{\rm IAS}^{\pm} &=& \sum_{i=1}^{A} \tau_\pm(i), \nonumber\\
F_{\rm GTR}^{\pm} &=& \sum_{i=1}^{A} [1 \otimes \overrightarrow{\sigma}(i)]_{J=1} \tau_\pm(i), \nonumber\\
F_{\rm SDR}^{\pm} &=& \sum_{i=1}^{A} [r_i Y_1(i) \otimes \overrightarrow{\sigma}(i)]_{J=(0,1,2)} \tau_\pm(i),
\eea
where 
$\sigma$ and $\tau$ are the Pauli matrices of spin and isospin degrees of freedom, respectively.
These operators are associated with the non-energy-weighted sum rules (NEWSR) for $S^- - S^+ = \sum_\nu B^-_\nu - \sum_\nu B^+_\nu$, accordingly:
\bea
  S^-_{\rm IAS} - S^+_{\rm IAS} &=& N - Z, \nonumber\\
  S^-_{\rm GTR} - S^+_{\rm GTR} &=& 3(N - Z), \nonumber\\
  S^-_{\rm SDR} - S^+_{\rm SDR} &=& \frac{9}{4\pi}\Bigr[N\langle r^2\rangle_n - Z\langle r^2\rangle_p\Bigr].  
\eea
Here we note that the spin-isospin response sum rules connect the two isospin channels with $T_{\pm}$ \cite{Osterfeld1992}. Indeed, they are usually tackled in a single calculation, which is in accordance with the general definition of the strength function (\ref{SF}).
The GTR NEWSR is known as the model-independent Ikeda sum rule. The SDR sum rule involves the root-mean-square radii of protons and neutrons, which makes this type of resonance relevant to measuring the neutron skin thickness \cite{Krasznahorkay1999, Yako2006}.
The $T_+$ excitations in neutron-rich nuclei are hindered by the Pauli principle so that the $S^-$ contributions dominate the NEWSRs in such systems.

The GTR is the most studied nuclear spin-isospin response because the transitions with $S = 1$, $J^\pi = 1^{+}$ are the major mechanism of the beta decay. By the nature of the transition operator, GTR probes the spin-orbit and isospin properties, so that the experimental data on the GTR provide information for constraining the respective terms in the effective interactions and EDFs.
A prominent example is one of the recently developed Skyrme effective interactions, SAMi \cite{RocaMaza2012}. It has implemented refined spin-isospin properties via the benchmark on the GTR peak energies. In the relativistic framework, it was demonstrated that an accurate description of GTR peak energies can be achieved on the (Q)RPA level by taking into account the Fock terms of the meson-exchange interactions \cite{LiangVanGiaiMeng2008, Niu2013, Niu2017}.


The (Q)RPA descriptions using effective interactions describe the major GTR peak quite reasonably \cite{Borzov:2003bb,Sarriguren2013,PaarNiksicVretenarEtAl2004a,LiangVanGiaiMeng2008, Niu2013, Niu2017}. Obtaining an adequate detailed strength distribution is impossible within (Q)RPA and requires the inclusion of the dynamical correlations. Another common problem is the reproduction of the total strength constrained by the Ikeda sum rule. Because of the model space limitations, in (Q)RPA it is exhausted within the relatively narrow energy region, which is the well-known quenching problem \cite{Osterfeld1992}. 
The situation is, thereby, similar to that with electromagnetic excitations: the effects beyond (Q)RPA are important for both gross and fine features of the GTR, and all the extensions beyond (Q)RPA improve the description of the GTR considerably.  Implementations with the qPVC kernels based on modern density functionals, both relativistic NL3 \cite{MarketinLitvinovaVretenarEtAl2012,
RobinLitvinova2016,RobinLitvinova2018,Robin2019} and non-relativistic Skyrme \cite{NiuNiuColoEtAl2015}, became available relatively recently, while the SRPA calculations for the GTR  were reported already in 1990 \cite{Drozdz:1990zz}. More recent SRPA with the Skyrme interaction also showed good performance \cite{Gambacurta2020}.  Since the low-energy part of the GTR spectrum in neutron-rich nuclei generates spontaneous beta decay, the description of beta decay rates is also improved by up to one or two orders of magnitude, when dynamical kernels are included, compared to those of (Q)RPA \cite{RobinLitvinova2016,LitvinovaRobinWibowo2020,Gambacurta2020}.

The latter is illustrated in Fig. \ref{GTR_Sn} showing the GTR in the neutron-rich tin isotopes $^{130,132,136}$Sn computed within the proton-neutron RQTBA (pnRQTBA) including the qPVC dynamical kernel, originally developed in Ref. \cite{RobinLitvinova2016}. The results are compared to the proton-neutron RQRPA (pnRQRPA) with the static kernel only. Both calculations employed the NL3 interaction for the static kernel, and the energy scales are relative to the parent nuclei. 
The qPVC kernel leads to a similar degree of fragmentation as in the electromagnetic response, and the dynamical correlations are more pronounced in nuclei with larger neutron excess. The fragmentation redistributes the strength at low energy, including the $Q_{\beta}$ window.  More strength in this energy window obtained in pnRQTBA leads to faster beta decay, as compared to pnRQRPA, and qPVC improves the agreement with data \cite{nndc}. The beta-decay half-lives extracted from the $Q_{\beta}$ energy interval of the GTR as described in Ref. \cite{RobinLitvinova2016} are shown in the right panel of Fig. \ref{GTR_Sn}. More examples of the GTR calculations can be found in Ref. \cite{LitvinovaBrownFangEtAl2014,RobinLitvinova2016} and the qPVC effects on the SDR were discussed in Ref. \cite{MarketinLitvinovaVretenarEtAl2012}. The pnRQTBA without superfluidity, i.e. pnRTBA, was extended to finite temperature with applications to the beta decay and electron capture in hot stellar environments \cite{LitvinovaRobinWibowo2020,Litvinova2021b}. 
\begin{figure}[t]
\begin{center}
\includegraphics[width=0.5\textwidth]{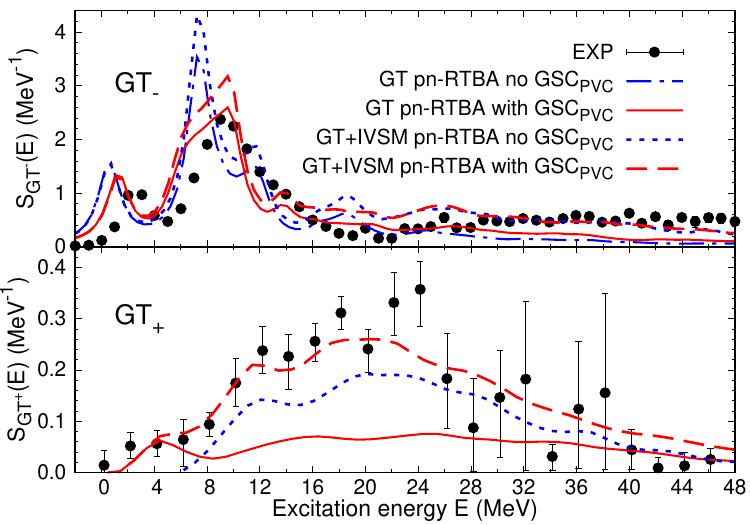}
\end{center}
\caption{The Gamow-Teller GT$_\pm$ strength functions for the ground-state-based transitions $^{90}\mbox{Zr} \rightarrow ^{90}\mbox{Nb}$ (top) and $^{90}\mbox{Zr} \rightarrow ^{90}\mbox{Y}$ (bottom).  The pure GT and combined GT+IVSM strengths obtained in pnRTBA without GSC$_{PVC}$ (dashed-dotted and dotted blue) and with GSC$_{PVC}$ (solid and dashed red) are shown, in comparison to the data of Refs. \cite{Yako2005,Wakasa1997}. 
The theoretical GT$_+$ and GT$_-$ strength distributions are smeared with a parameter $\Delta = 2$ and $1$ MeV, respectively, to match the experimental energy resolutions. The figure is adapted from Ref. \cite{Robin2019}.}
\label{f:90Zr}%
\end{figure}

A more advanced calculation with the non-superfluid PVC kernel included the ground state correlations (GSC) caused by PVC (GSC$_{\text{PVC}}$) beyond the leading-order resonant NFT terms. Such correlations were introduced and implemented,e.g., in Refs. \cite{KamerdzhievTertychnyiTselyaev1997,KamerdzhievSpethTertychny2004}, where their role was found significant, particularly for the spin-flip magnetic dipole excitations. Since GTR also involves the spin-flip, an important contribution from the GSC$_{\text{PVC}}$ in its formation is expected. The significance of the GSC$_{\text{PVC}}$ for the GT$_+$ branch in neutron-rich nuclei was hypothesized based on the potential of these correlations to unblock the GT$_+$ transitions involving configurations more complex than the $ph$ ones. A comprehensive study of this case was performed for $^{90}$Zr in \cite{Robin2019} exemplified in Fig. \ref{f:90Zr}, in comparison to data of Refs. \cite{Yako2005,Wakasa1997}. 
In the GT$_-$ branch, the inclusion of the PVC effects within the resonant pnRTBA leads to the fragmentation and broadening of the strength distribution with respect to the pnRRPA as expected and pointed out in the previous cases. 
The GT$_+$ transitions from particle to hole states are unlocked by the GSC of RPA (GSC$_{\text{RPA}}$), but the $ph$ transitions appear only above 7 MeV with insignificant probabilities.
The PVC in the pnRTBA with only the leading resonant NFT forward-going diagrams essentially does not change this result.
The inclusion of the GSC$_{\text{PVC}}$ with the backward-going PVC contributions, however, has a significant effect on the GT$_+$ strength. These correlations induce fractional occupancies of the single-particle orbitals in the parent nucleus, which unlocks transitions between the states on the same side of the Fermi surface. As an example, the peak at $4.5$ MeV is dominated by the $\pi 1g_{9/2}\to \nu1g_{7/2}$ and $\pi 2p_{3/2} \to \nu 2p_{1/2}$ transitions, with the notable absolute values of the transition densities of $0.347$ and $0.182$, respectively.
In the GT$_-$ branch, pnRTBA with GSC$_{\text{PVC}}$ reaches a good agreement with the data up to $\sim 25$ MeV, except for the position of the low-lying state showing some mismatch.
In the GT$_+$ channel, the PVC-induced GSCs are solely responsible for the appearance of both the low-energy peak at $4$ MeV and the higher-energy broadly distributed strength up to $\sim 50$ MeV. Above the low-energy peak, the pnRTBA GT$_+$ strength alone underestimates the experimentally observed one. However, it is known that the isovector spin-monopole (IVSM) mode becomes important at high excitation energies.  In particular, the data of Refs. \cite{Yako2005,Wakasa1997} contain the IVSM contribution, which could not be disentangled from the GT transitions, therefore, the IVSM was also added to the theory.
The IVSM mode was obtained as the response to the operator $F_{\text{IVSM}}^{\pm} = \sum_{i} r^2(i) {\overrightarrow\Sigma}(i) \tau_{\pm}(i)$, which should be superposed with the GT one, for instance, following the procedure of Ref. \cite{Terasaki2018}. As a result, the mixed operator $F_{\alpha}^{\pm} = \sum_{i} [1 + \alpha r^2(i)] {\overrightarrow\Sigma}(i) \tau_{\pm}(i)$, with the parameter $\alpha$ adjusted to the magnitude of the theoretical low-energy GT strength was adopted in the calculations. 
The values  $\alpha = 9.1 \times 10^{-3}$ and $\alpha = 7.5 \times10^{-3}$ fm$^{-2}$, respectively, were used for the GT$_+$ and GT$_-$ channels.
The resulting strength delivers an improved description of the experiment also above 25-30 MeV in both GT$_{\pm}$ branches, thus further justifying the importance of both the GSC$_{\text{PVC}}$ and the IVSM contributions for the data interpretation \cite{Robin2019}.

\section{Summary and outlook}

In these lectures, I discussed fermionic in-medium equations of motion and their applications to nuclear structure with a major focus on the nuclear response. We have seen that the EOMs for the two-time propagators provide a formally complete theory to describe nuclear spectra, however, they admit only approximate solutions.
Starting from the most general many-body Hamiltonian, we constructed a model-independent theoretical framework for the lowest-rank in-medium fermionic propagators. Approximations, most relevant in the intermediate and strong coupling regimes, to the commonly accessible observables were derived and discussed. The new scale with the order parameter associated with emergent collective degrees of freedom was used to advance the dynamical interaction kernels in a systematically improvable way.

The EOM for the one-fermion propagator was worked out in a single-particle basis, first in the absence of superfluidity. Subsequently, we added superfluid pairing correlations and transformed the single-particle sector of the theory to the HFB basis, relaxing the particle number constraint. This representation enabled scaling the computational effort down considerably and paved the way for the superfluid response theory, which was built in the HFB basis from the start. In particular, the unification of particles and holes in the quasiparticle concept established the framework to unify the particle-hole and particle-particle channels in the response theory.
The dynamical interaction kernels appearing in the integral parts of the EOMs in terms of higher-rank propagators were discussed in detail as they represent the major obstacle to accurate solutions of the nuclear many-body problem. The latter propagators are approximated by factorizations into the possible products of CFs. For practical applications to nuclear structure,  introducing a truncation of the many-body problem on the two-body level is possible by retaining two-fermion and one-fermion CFs. The presence of the two-body CFs is found to be the minimal requirement for keeping the leading effects of emergent collectivity. Then, we discussed how, by gradually relaxing correlations, the theory with factorized dynamical kernels can be reduced to further approximations.

The practical inclusion of the quasiparticle-vibration coupling, or qPVC, in the dynamical kernels of the response theory was then discussed in the context of numerical implementations for nuclear spectral calculations. Selected results obtained in the RNFT framework were presented for the dipole and Gamow-Teller responses of medium-mass nuclei. The leading-order qPVC effects are shown to considerably modify 
the spectral strength distributions, obtained in the relativistic QRPA, causing fragmentation of the QRPA modes and shifting the positions of the major peaks. An example of a higher configuration complexity up to $2q\otimes 2phonon$ was considered for the electromagnetic dipole strength distribution in medium-light nuclei. Overall, the increase of configuration complexity brings the theoretical results in better agreement with the data \cite{LitvinovaSchuck2019,Novak2024} and the resulting strength functions show saturation of its general features with the increase of configuration complexity. Another type of complex correlations, namely the qPVC-associated ground state correlations, was found to play a sizeable role in the nuclear response induced by weak interactions. The case of GT$_+$ response of a neutron-rich $^{90}$Zr was brought up as an example of this kind.


Since the response functions of correlated quantum systems are deeply linked to the in-medium propagators, their theory is transferrable across energy scales, and applications connect various disciplines.  Nuclear response, in particular, finds astrophysical 
applications, where accurate spectral computation is required. The studies of rapid neutron capture process nucleosynthesis, or r-process,  in neutron star mergers, core-collapse supernovae, and ultra-high energy cosmic rays need the nuclear structure input at the extremes of energy, mass, isospin, and temperature. 
Nuclear electric and magnetic dipole, Gamow-Teller, and spin-dipole response functions provide the reaction rates for the radiative neutron capture 
$(n,\gamma)$, electron capture, $\beta$ decay, and $\beta$-delayed neutron emission under astrophysical conditions. These reaction rates are sensitive to the fine details of the theoretical strength functions and are needed for a massive amount of atomic nuclei which are not measurable in the laboratory. The accurate knowledge of soft modes of response, i.e., the low-energy parts of the strength distributions is especially important. The low-lying dipole strength has been studied very intensively during the past decades and is associated with neutron skin oscillations against the nearly isospin-saturated core. Such oscillations are most pronounced in the neutron-rich nuclei constituting the r-process path in the nuclear landscape and form the pygmy dipole resonance, which affects the $(n,\gamma)$ rates considerably \cite{LitvinovaRingTselyaevEtAl2009,LitvinovaLoensLangankeEtAl2009,
SavranAumannZilges2013,PaarVretenarKhanEtAl2007}. The low-energy shoulders of the GTR and SDR strength functions provide information about the beta decay and electron capture rates \cite{ NiksicMarketinVretenarEtAl2005a,Niu2013,Mustonen2016,Dzhioev2020}.  These weak reaction rates are affected tremendously by the correlations beyond QRPA \cite{Niu2013,NiuNiuColoEtAl2015,RobinLitvinova2016,RobinLitvinova2018,LitvinovaRobinWibowo2020,Litvinova2021b}, however,
the simplistic QRPA reaction rates are still employed in most astrophysical simulations \cite{ArnouldGorielyTakahashi2007,MumpowerSurmanMcLaughlinEtAl2016,Langanke2021,Cowan2021}. 
The deficiencies of the nuclear spectral input are strongly amplified in the modeling of stellar environments so that a high-quality theory and its implementations have the potential to considerably advance the simulations of kilonovae and supernovae.

The nuclear structure applications to the searches for new physics beyond the Standard Model, such as the neutrinoless double beta decay and the electric dipole moment also depend sensitively on the quality of the nuclear spectral description. These phenomena result from a delicate interplay of complex effects of the correlated nuclear media and require computation of consistency and accuracy going beyond the current state-of-the-art of the nuclear response theory.
This further stresses the importance of new developments in quantum many-body theory compatible with special relativity and rooted in particle physics, first of all, in the direction of complex dynamical configurations. Reconciling consistently the static and dynamical kernels of the two-fermion EOMs is needed to link the NN potentials with the non-perturbative in-medium physics dominated by the emergent scales of collective degrees of freedom.

\section*{Acknowledgement}
This work was supported by the GANIL Visitor Program, US-NSF Grant PHY-2209376, and US-NSF Career Grant PHY-1654379.
%
\bibliography{Bibliography_Jun2024}
\end{document}